\documentclass[useAMS,usegraphicx,usenatbib]{mn2e}
\usepackage{aas_macros}
\usepackage[a4paper,centering, totalwidth=520pt, totalheight=700pt]{geometry}
\usepackage[fleqn]{amsmath} 
\usepackage{graphicx}
\usepackage{amssymb}
\usepackage{amsmath}
\usepackage{enumerate}
\usepackage{microtype}
\usepackage[usenames,dvipsnames]{color}

\newcommand{\Mass}{\ensuremath{ h^{-1} M_{\odot}}}

\title[Refined phase-space elements]
{An adaptively refined phase-space element method for cosmological simulations and collisionless dynamics}
\begin{document}

\author[O. Hahn \& R.~E. Angulo]{
Oliver Hahn\thanks{Email:
    hahn@phys.ethz.ch}$^{1}$ and Raul E. Angulo\thanks{Email:
    rangulo@cefca.es}$^{2}$ \\
  $^{1}$Department of Physics, ETH Zurich, CH-8093 Z\"urich,
  Switzerland \\ 
  $^{2}$Centro de Estudios de F\'isica del Cosmos de Arag\'on, Plaza San Juan 1,  Planta-2, 44001, Teruel, Spain.
}

\date{submitted to MNRAS Jan. 8, 2015}

\maketitle

\begin{abstract}
$N$-body simulations are essential for understanding the formation and evolution of structure in the Universe. However, the discrete nature of these simulations affects their accuracy when modelling collisionless systems. We introduce a new approach to simulate the gravitational evolution of cold collisionless fluids by solving the Vlasov-Poisson equations in terms of adaptively refineable ``Lagrangian phase space elements''. These geometrical elements are piecewise smooth maps between Lagrangian space and Eulerian phase space and approximate the continuum structure of the distribution function. They allow for dynamical adaptive splitting to accurately follow the evolution even in regions of very strong mixing.

\noindent We discuss in detail various one-, two- and three-dimensional test problems to demonstrate the performance of our method. Its advantages compared to $N$-body algorithms are: i) explicit tracking of the fine-grained distribution function, ii) natural representation of caustics, iii) intrinsically smooth gravitational potential fields, thus iv) eliminating the need for any type of ad-hoc force softening.

\noindent We show the potential of our method by simulating structure formation in a warm dark matter scenario. We discuss how spurious collisionality and large-scale discreteness noise of $N$-body methods are both strongly suppressed, which eliminates the artificial fragmentation of filaments. 

\noindent Therefore, we argue that our new approach improves on the $N$-body method when simulating self-gravitating cold and collisionless fluids, and is the first method that allows to explicitly follow the fine-grained evolution in six-dimensional phase space.
\end{abstract}



\section{Introduction}

Numerical simulations lie at the very heart of contemporary cosmology. They are
the only method that can accurately follow the growth of small primordial
density fluctuations into the highly nonlinear objects that populate the
low-redshift Universe \citep[e.g.][]{Davis1985, Efstathiou1985,
Bertschinger1998,Springel2005a, Angulo2012a}. As such, they have proven an
indispensable tool in the formulation of our theory of cosmological structure
formation and in the validation of the $\Lambda$CDM model.

Since most of the mass in the Universe appears to be in the form of dark
matter (DM; a fundamental particle with a negligible non-gravitational
interaction cross-section with both itself and baryonic matter),
numerical simulations that only follow
gravitational forces were the natural first tool employed by pioneer numerical
cosmologists. Since the 1970s, these simulations have progressively increased
their scope and accuracy, nowadays spanning a huge dynamic range.
State-of-the-art simulations employ trillions of bodies to describe volumes
comparable to the observable Universe, while resolving the collapsed DM
structures that could host the faintest galaxies 
\cite[see e.g.][for recent
examples]{Heitmann2014b,Skillman2014,Ishiyama2014}.

A milestone in the history of gravity-only simulations was the establishment of
a universal form for the density profile of collapsed dark matter haloes
\citep{Navarro1996,Navarro1997}. Other important results were the accurate
characterisation of the hierarchy of nonlinear correlation functions, of the
cosmic velocity field, precise estimates of nonlinear effects on the BAO peak
and the universal form for the abundance of dark matter haloes, which secure
their use for cosmological parameter estimation \cite[see the reviews of][and
references therein]{Kuhlen2012,FrenkWhite2012}.

The central role of simulations in cosmology is in part due to the apparent
robustness of the methods employed. Their results seem to hold even when
extremely different time, mass and force resolution are used.  For instance,
regarding the functional form and universality of halo density profiles,
state-of-the-art simulations show a remarkable agreement with those carried out
$20$ years ago, despite employing $5$ orders of magnitude more particles, and
$100$ times better force resolution \citep[e.g.][]{Diemand2008,
Springel2008,Gao2012}. It is worth noting that, despite some recent hints
\citep{Pontzen2013,Ludlow2013,Ludlow2014}, the fundamental origin of the halo
density profile is still unknown, and even a purely numerical one has been
suggested \citep[e.g.][]{Baushev2015}. 


\subsection{The Collisionless Boltzmann equation}

Dark matter (DM) in the Universe can be considered as a collisionless fluid.
Even in the case of a (heavy) 100~GeV DM particle, the mean comoving 
number density of particles is $\sim10^{48}\,{\rm pc}^{-3}$. This obviously implies that dark matter,
just like baryons, can be treated as a fluid on cosmological and astrophysical
scales, where the microphysical particle character is irrelevant. Additionally, DM appears to have an extremely weak
interaction cross-section with itself or with ordinary matter. Therefore, the
DM fluid can be regarded as collisionless. These properties hold for most
particle dark matter candidates, even for warm ones such as e.g. sterile
neutrinos. 

A self-gravitating collisionless fluid is fully described by the evolution of
its six-dimensional phase-space density $f(\mathbf{x},\mathbf{p},t)$, which is governed by
the collisionless Boltzmann equation (CBE):

\begin{equation}
0=\frac{{\rm d} f(\mathbf{x},\mathbf{p},t)}{{\rm d}t} = \frac{\partial f}{\partial t} + \frac{\mathbf{p}}{ma^2}\cdot\boldsymbol{\nabla}_{x}f - m\boldsymbol{\nabla}_{x}\phi\cdot\boldsymbol{\nabla}_{p} f,
\label{eq:boltzmann}
\end{equation}

\noindent supplemented with Poisson's equation for the Newtonian gravitational potential $\phi$

\begin{equation}
\boldsymbol{\nabla}_{x}^{2} \phi = \frac{4\pi G m_p}{a}\left[ \left(\int {\rm d}^3p \,  f\right) -  \bar{n} \right],
\label{eq:Poisson}
\end{equation}

\noindent where $m_p$ is the (microphysical) particle mass, $\bar{n}$ is the mean number density
of dark matter particles in the Universe, $G$ is the gravitational constant and
$a$ is the cosmological scale factor that itself obeys the first Friedmann
equation. The system of the CBE with a self-coupling described by Poisson's
equation is commonly termed the Vlasov-Poisson system of equations. While we focus here
on modelling the DM dynamics, we note that Vlasov-Poisson describes all collisionless fluids 
with a $1/r$ potential and thus the discussion presented in this paper also applies to those cases.

One can consider two limits for the initial distribution function,
$f(\mathbf{x},\mathbf{p},t=0)$. In the {\em hot} limit, the problem is manifestly
{\em six-dimensional}, where the thermal width gives a natural scale in
velocity space that one can aim to resolve. Quite in contrast, in the {\em
cold} limit, i.e. when the thermal particle velocity dispersion is much smaller than the
bulk velocities arising from gravitational instability (which is an excellent
approximation for both cold and warm DM candidates\footnote{In fact, for a $1\,{\rm keV}$ particle, the RMS velocity
is only around $\sim0.4\,{\rm km/s}$ at $z=9$ \citep{Bode2001} when structure formation starts to become
nonlinear. This velocity further decays and is orders of magnitude smaller than the velocities arising from collapse. 
The signature of the thermal velocity {\em is} the truncation of the spectrum, which represents the largest Jeans mass before
the particles start to cool. At later times, the particles are never substantially heated again since they remain locally (on their stream) roughly at the cosmic 
mean density even inside collapsed structures \citep[cf.][]{Vogelsberger2011}. In very good approximation, it thus suffices to consider all 
(non-relativistic) dark matter particles as a cold fluid.}), the problem is only
{\em three-dimensional}:  the thermal width is zero and thus the fluid
occupies only a three-dimensional submanifold of phase space \citep[cf.
e.g.][]{Zeldovich1970,Arnold1982}. This means that the distribution function is a smooth (i.e. differentiable) 
three-dimensional hypersurface without holes in six-dimensional phase-space. The Vlasov-Poisson
equation guarantees that it remains smooth and that no holes will emerge in phase space during its evolution. 
In particular, one can parameterise this
submanifold $\mathcal{Q}$ in terms of Lagrangian coordinates $\mathbf{q}$ (so
that it is customary to call it the ``Lagrangian manifold'') and express the
mapping between Lagrangian and Eulerian phase space as

\begin{equation}
\mathcal{Q}\subseteq\mathbb{R}^3\to\mathbb{R}^3\otimes\mathbb{R}^3:\quad\mathbf{q}\mapsto\left(\mathbf{x_q},\mathbf{v_q}\right), \label{eq:the_map}
\end{equation}

\noindent whose evolution describes the full evolution of the fluid in both the
monokinetic (single-stream) and the multi-stream regime. Despite the cold
fine-grained distribution function, this three dimensional hyper-surface can
become ``dense'' in phase space  through gravitational evolution and mixing,
which means that it will cover more and more of phase space over time and
approach a state close to a macroscopically ``hot'' system. Once the phase
space is densely covered (meaning that the distance in phase space between
different regions of the sheet becomes small compared to the extent of the
system), one can expect that the system does not change any more on time scales
on which particles move, which means that it will approach ergodicity in those
regions. 

Unfortunately, the solution of the set of equations in the hot limit without
further approximations is extremely expensive computationally, and an explicit
6D solution has been obtained only for a handful of simplified cases
\citep[e.g.][using a $64^6$ grid]{Yoshikawa2012}. The situation is in some
sense even worse in the cold limit, since both spatial and velocity resolution
achievable with such a coarse-grained approach are completely insufficient to
resolve the dynamics of a cold system without being dominated by diffusion in
phase space.  Additionally, because of the collisionless nature of the DM
fluid, Boltzmann's H-theorem does not apply (in the sense that it is not to be
expected that the system approaches a maximum entropy state described
by a single ``temperature'' on short-enough time scales) and hence the CBE cannot be
replaced by a low-order moment expansion in terms of fluid equations with a
simple closure condition such as an effective pressure or effective
viscosity \citep[although such approaches are applied on mildly non-linear
scales, e.g.][but the effective fluid properties have not been determined from
first principles, and are provided by virialisation on smaller scales rather
than a conventional thermalisation]{Carrasco2012,Hertzberg2014}.

To make progress, it has thus been customary to postpone the goal of following
the exact evolution of the fine-grained distribution function and instead to
solve the Vlasov-Poisson system using a Monte-Carlo approach. The idea is to
sample the distribution function at $N$ discrete locations $\mathbf{q}_i$
($i=1\dots N$) with massive bodies (in what can be thought of as a
``coarse-graining'' of the initial conditions over macroscopic volume elements $\delta V$
containing a mass $m$ of microscopic particles, i.e. $m = m_p\, \bar{n}\, \delta V$):

\begin{equation}
f_N(\mathbf{x},\mathbf{p},t) = \sum_{i=1}^N \delta_D\left(\mathbf{x}-\mathbf{x}_i(t)\right)\,\delta_D\left(\mathbf{p}-\mathbf{p}_i(t)\right),
\label{eq:pointwise_psdf}
\end{equation}

\noindent for which the Vlasov-Poisson system then reduces to the equations of
motion for a Hamiltonian $N$-body system, i.e. the phase space density $f_N$ is
trivially conserved along the characteristics $\mathbf{x}_i(t)$,
$\mathbf{p}_i(t)$ defined by:

\begin{equation}
\dot{\mathbf{x}}_i = \frac{1}{m\,a^2}\mathbf{p}_i \quad\textrm{and}\quad
\dot{\mathbf{p}}_i = -m\left.\boldsymbol{\nabla}_x \phi\right|_{\mathbf{x}_i}. 
\label{eq:equations_of_motion}
\end{equation}
The equations of motion are thus easy to solve if the gravitational potential is known.

Inserting the distribution function (\ref{eq:pointwise_psdf}) into Poisson's equation (\ref{eq:Poisson}),
and convolving with the Green's function of the Laplacian $G(r)=-1/4\pi r$, the potential is given by
the Newtonian potential of $N$ point masses. The initial coarse graining however implied that the
mass is not concentrated in one point, but spread out over the volume $\delta V$, and so the unbounded
two-body force is certainly a bad approximation to the continuous collisionless system. The evolution of
the volume $\delta V$ is non-trivial and so an approximation is made by replacing the point mass 
potential e.g. with a Plummer softened potential
\begin{equation}
\phi(\mathbf{x}) =  - \sum^N_{j=1} \frac{G m}{ \sqrt{ (\mathbf{x} - \mathbf{x}_j)^2 + \epsilon^2}},
\end{equation}
\noindent where $\epsilon$ is the softening
length (which may evolve over time). In principle initially $\epsilon \sim \delta V^{1/3}$ 
is of the order of the mean particle
separation and becomes smaller in regions of collapse\footnote{In fact, $\delta V$ represents the
phase space volume associated with a particle and will remain of the order of the mean particle
separation before shell crossing. After shell-crossing in regions of multi-streaming, it is still on the order of the
mean separation, but the mean separation of particles on the same stream, not the mean separation
after projection to configuration space as is done when adaptive softening is used (either
by adaptive softening lengths or in AMR). It is thus hopeless to estimate $\epsilon$ more accurately
using only configuration space properties.}. Since its evolution is not followed, 
it is usually treated as a somewhat arbitrary parameter to soften the force field on a
small fraction of the mean particle separation scale that seeks to suppress
the discrete nature of the system while allowing for the highest force ``resolution''
possible. The validity of this approach is then typically inferred from convergence studies
in which $N$ and $\epsilon$ are varied. In fact, the method we present in this paper can be
viewed as following the exact evolution of the $\delta V$ volume elements over time.

The evolution of the $N$-body system is thus assumed to be analogous to the
evolution of a macroscopic group of microphysical dark matter particles, and therefore, it
is assumed that it is an actual solution to the CBE. This is clearly true {\it only}
when $N$ goes to infinity and the force softening $\epsilon$ goes to zero. For
a finite number of particles, there are collision terms that would affect the
details of dynamics such as chaotic and phase mixing, Landau damping and
two-body interactions among others. Furthermore, it is not possible to
guarantee an upper bound to the error introduced. This can lead to solutions
dominated by numerical artefacts.

Over the years, many authors have pointed out cosmological set-ups in which the
discrete and collisional nature of the $N$-body solution can be appreciated
\citep[e.g.][]{Centrella1983,Centrella1988, Melott1989, Diemand2004,
Melott1997, Splinter1998, Wang2007a, Melott2007, Marcos2008}. There are two
situations worth mentioning.  The first is simulations that feature a
small-scale cut-off in their primordial density power spectrum and in which the
$N$-body method produces a large number of spurious clumps that resemble dark
matter haloes (usually known as ``artificial filament fragmentation''). Another
example is when the gravitational interaction of two fluids that
start with a different power spectrum is followed, e.g. baryons and dark matter in the
early Universe. In this case, two-body interactions make the fluids
collisional resulting in an incorrect evolution of their relative spatial
distribution \citep{Yoshida2003,Angulo2013}. In all these cases problems are 
most severe if $\epsilon$ is smaller than the mean particle separation 
(which is however clearly the case in all production run $N$-body simulations where 
typically $\epsilon$ is $1/20$ to $1/60$ of the mean particle separation).

It is not yet clear how the problems described above affect standard
single-fluid CDM simulations. However, there is a reasonable amount of concern
about the correctness and accuracy of $N$-body simulations. The importance that
numerical simulations have for modern cosmology certainly warrants a level of
skepticism and motivates the search for alternative methods that do not
suffer from the problems of the $N$-body method.


\subsection{A phase-space element method}

In order to overcome the limitations of the traditional $N$-body approach,
various improvements and alternative schemes have been proposed over the years. 

Most $N$-body simulations employ a softening scale that is either constant or
varies only over time. Several attempts have been made using spatially varying
adaptive softening \citep[e.g.][]{Bagla2009,Iannuzzi2011}, which suppresses
two-body interactions more efficiently. Adaptive softening is, however,
performed isotropically, which implies that it is not expected to improve
dramatically the situation in regions of anisotropic collapse (which in
cosmological simulation is basically everywhere except inside of haloes). 

Among the methodological alternatives to the $N$-body method are e.g.  a
reformulation of Vlasov-Poisson as a Schr\"odinger equation \citep[e.g.][for
classical and more recent examples]{Widrow1993,Schaller2014,Uhlemann2014}, the
1D waterbag method \citep[e.g.][]{Colombi2014,Colombi2015}, as well as the
Lagrangian tessellation method of  \cite{Hahn2013}, hereafter HAK13.

The method proposed in HAK13 (referred to as ``TetPM'' by the authors) attempts
to retain the continuous nature of the distribution function $f$ by performing a
Delaunay tessellation of the $N$-body particles in Lagrangian space (based on
the Lagrangian tessellation idea of \citealt{Abel2012} and
\citealt{Shandarin2012}). HAK13 demonstrated that by assigning the mass to
tetrahedral volume elements, whose vertices are free-falling particles, anisotropic
deformation can be followed much more accurately and the
obvious discreteness effects of the $N$-body method, such as two-body collisions and
spurious fragmentation, could be overcome. However, a serious shortcoming
of the method was also noted by HAK13: approximating the manifold by
tetrahedra is accurate in the mildly nonlinear regime, but not
in strongly non-linear stages. In this regime, the volume of $f$ grows faster than what
can be followed by the motion of the tetrahedra. This leads to a violation of
the Hamiltonian structure of the equations, and thus it implies a lack of energy
conservation. This can manifest itself in an overestimation of the density in
the centre of haloes (HAK13), and it also affects the location of material tidally
disrupted from infalling dark matter substructure \citep{Angulo2014a}.

In this article, we present a general class of {\em "Lagrangian phase space
element methods"} to solve the Vlasov-Poisson system with cold initial conditions. 
We will show that the approach of HAK13 is
naturally included as the lowest order method that provides a
continuous approximation to the distribution function.  When higher order
elements are employed, the evolution of phase-space elements can be captured
much more accurately, and caustics can be explicitly represented. Furthermore,
we will demonstrate that an adaptive refinement of phase space elements allows
to fully track their distortion, even during late highly nonlinear stages. This
guarantees energy conservation and a direct control of errors. Therefore, we
will argue that our method offers an interesting alternative to solve the CBE
while fully resolving the evolution of the full phase-space distribution function.
The possibility to have the full fine-grained distribution function at hand offers
exciting possibilities to study the behaviour of self-gravitating collisionless systems
in much more detail than can be extracted from $N$-body data.

Our paper is structured as follows. In Section~\ref{sec:method}, we introduce
the phase space element method, discuss how we implemented the
projection to configuration space for the mass deposit, discuss discretisation
errors, and how they can be limited by the adaptive dynamic refinement
technique that we also present. After having introduced all methodology, in
Sections~\ref{sec:testing} and \ref{sec:testing_gravity}, we validate the
performance of our method in various one-, two-, and three-dimensional test
problems first without and then with self-gravity. Finally, we apply our method
to a cosmological simulation with a cut-off in the initial spectrum in
Section~\ref{sec:cosmo}, and demonstrate the advantages over previous
approaches. We discuss our results and conclude in
Section~\ref{sec:conclusions}.



\section{Method}
\label{sec:method}

In what follows, we will present the theoretical foundations, as well as a
practical implementation, of a new general class of numerical methods to solve
the CBE referred to as ``{\em Lagrangian phase space elements}''. 

We start in \S2.1 by introducing the fundamental unit of our method: piecewise
maps between Lagrangian space and Eulerian phase space. Then we discuss
how this gives rise to piecewise density fields that are defined everywhere in
space and can be made arbitrarily regular, but at the same time can contain
caustics of various types. In \S2.2 we focus on the errors associated to our
scheme and how they propagate over time. In \S2.3 we exploit this and show how
the Lagrangian phase space elements can be adaptively split to track the full
phase-space evolution of a fluid to a given accuracy. In \S2.4 we discuss how
our method allows an accurate and efficient calculation of the evolution of a
self-gravitating fluid. In the final section, \S2.5, we further discuss  the
implementation of these ideas and parallelization strategies for an efficient
execution in distributed-memory computers.


\subsection{Lagrangian phase space elements}

\begin{figure}
\begin{center}
\includegraphics[width=0.49\textwidth]{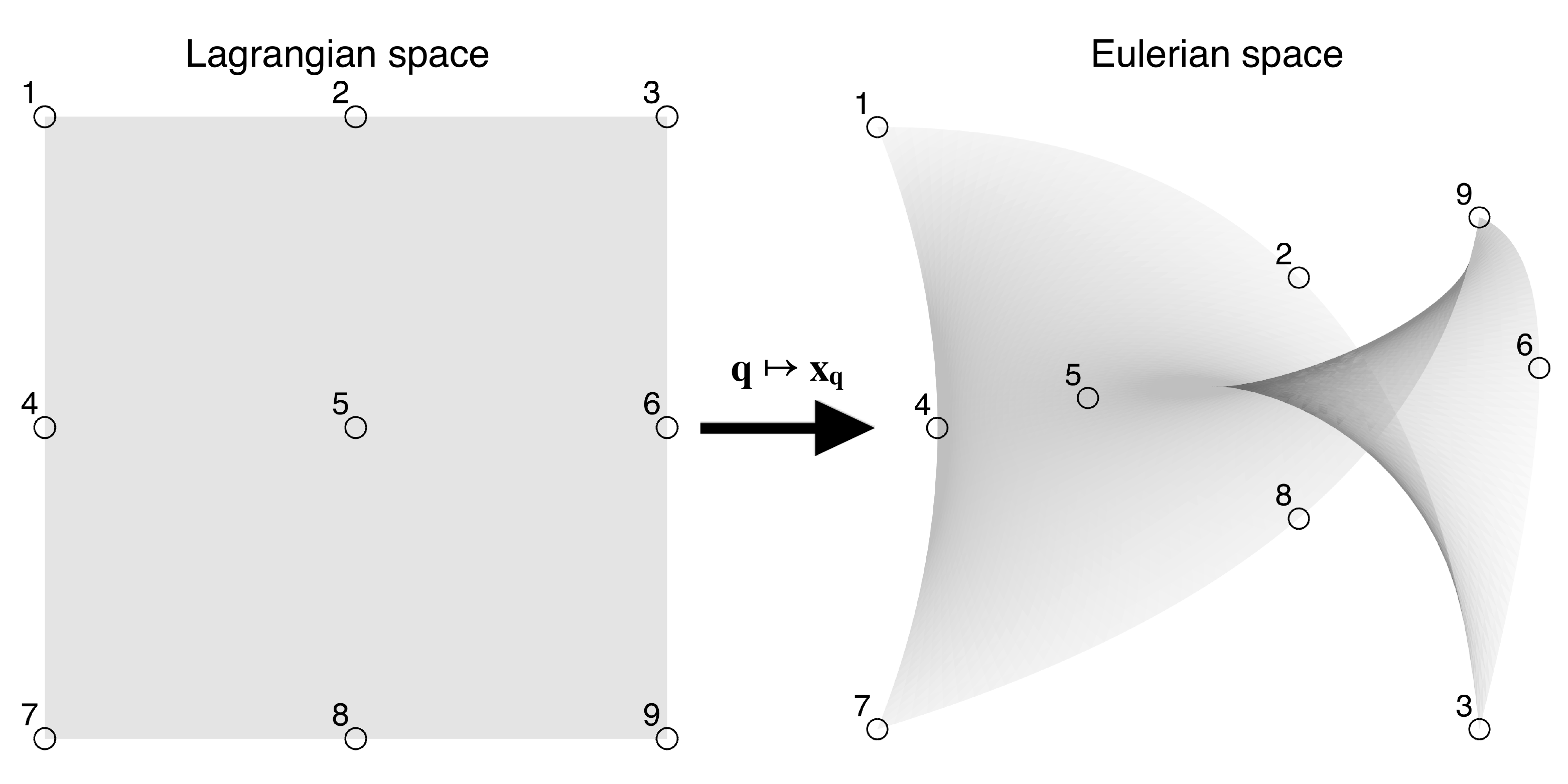}
\end{center}
\caption{
\label{fig:lagrange_mapping}Mapping between Lagrangian space and Eulerian configuration space. The mapping can become multi-valued in projection to configuration space and can have singular points and curves. Here, as an example, the mapping of nine points forming a square in Lagrangian space are shown under a biquadratic  map to phase-space and then projected into configuration space.
}
\end{figure}

\subsubsection{The Lagrangian manifold}
Let us begin by considering in more detail the evolution of the neighbourhood of a point with Lagrangian coordinate $\mathbf{q}\in\mathcal{Q}$, which maps to Eulerian phase space as $\mathbf{q}\mapsto\left(\mathbf{x_q}(t),\mathbf{v_q}(t)\right)$ at some time $t$. An illustration for such a mapping in two dimensions is given in Fig.~\ref{fig:lagrange_mapping}. Shown is a piece of Lagrangian space, a square in this case, on which nine points are put down on a regular lattice. In Lagrangian space the density is uniform. This square is now mapped into four-dimensional phase space (where it is still a two-dimensional surface) and then projected back into two-dimensional configuration space. In the case shown, the map was given by some bi-quadratic polynomial. In general, the projection can lead to several points in Lagrangian space mapping to the same point in Eulerian configuration space\footnote{The map to configuration space is thus in general surjective but not injective, while the one to phase space is bijective.}. The projection becomes singular -- corresponding to caustic subsets -- where derivatives of the map to configuration space vanish (i.e. where $\boldsymbol{\nabla}_\mathbf{q}\otimes\mathbf{x_q}$ has at least one vanishing eigenvalue). The singularities arising from such projections have been studied in catastrophe theory in great detail \citep[e.g.][for caustics of the Zel'dovich map in two and three dimensions]{Arnold1982,Hidding2014}. In what follows, we describe a procedure to produce a covering of the Lagrangian domain with such local maps.

The differential nature of the cold distribution function guarantees that the space tangent to the sheet at $\mathbf{q}$ in phase space is uniquely defined as,
\begin{equation}
T_\mathbf{q}\mathcal{Q}: \mathbf{q} \mapsto \left( \boldsymbol{\nabla}_\mathbf{q} \otimes\mathbf{x_q}, \boldsymbol{\nabla}_\mathbf{q} \otimes\mathbf{v_q} \right).
\end{equation}
This tangent space evolves under its own set of equations, sometimes called the ``geodesic deviation equation'' (GDE) because of the analogy to geodesics in curved spaces \citep{Vogelsberger2008,White2009,Vogelsberger2011}, and the tetrahedral approximation is a secant approximation to this tangent space\footnote{For tetrahedra, the $\boldsymbol{\nabla}_q$ are just given through the three directional finite differences that can be computed on a tetrahedron.}. Since the GDE gives us only infinitesimal information around a point, and the tetrahedral secants are very crude, a better local description of $f$ in the neighbourhood of $\mathbf{q}'$ can be achieved by considering the evolution of a higher order expansion of $(\mathbf{x_q},\mathbf{v_q})$ in the neighbourhood of $\mathbf{q}'$. 

\subsubsection{Lagrangian elements}
The central idea of our approach is to consider a decomposition of the Lagrangian manifold into a finite number of Lagrangian volume elements $\delta V$. In the TetPM approach, these volume elements were tetrahedra, but in principle any volume decomposition is possible. In what follows we will use cubical volume elements to decompose Lagrangian space. {\em Each element carries constant mass and defines a local map between Lagrangian space and Eulerian phase space}. In particular, we will approximate this map using multi-variate polynomials in what follows. We will show that the coefficients of these polynomials, each defined on its Lagrangian cubical element, can be represented by a number of supporting points (or flow tracer particles). 

For a three-dimensional Lagrangian manifold, we will thus consider the space of 3-variate polynomials $P_k$ of order $k$ in $\mathbf{q}=(q_0,q_1,q_2)$, i.e. the set
\begin{equation}
P_k = \{ \pi(\mathbf{q}) \,\, | \,\, \pi(\mathbf{q})  = \sum_{\alpha,\beta,\gamma=0}^k a_{\alpha\beta\gamma}\, q_0^\alpha q_1^\beta q_2^\gamma\},
\end{equation}
and we note that ${\rm dim}\,P_k = (k+1)^3$. The Lagrangian environment $\delta V$ of our point $\mathbf{q}'$ shall be described by a unit cube $K_3=[0,1]^3$ centred on $\mathbf{q}'$. Matching ${\rm dim}\,P_k$, we can choose the following subset of points  $\Sigma_k\subset K_3$ on a regular lattice
\begin{equation}
\Sigma_k \equiv \left\{ \left. \left(\frac{i_0}{k},\frac{i_1}{k},\frac{i_2}{k}\right)\subset K_3 \,\,\right|\,\, i_j\in\left\{0,1,\dots,k\right\} \right\},
\label{eq:supporting_points}
\end{equation}
so that the $(k+1)^3$ coefficients $a_{\alpha\beta\gamma}$ are fully determined if $\pi(\mathbf{b})$ is known for {\em all} $(k+1)^3$ elements $\mathbf{b}\in\Sigma_k$. We also say that the element has $(k+1)^3$ {\em degrees of freedom}. Note that the requirement that $\Sigma_k$ be a regular lattice is not a necessity but a mere convenience when calculating the polynomial coefficients. Furthermore, the lattice is ``force-free'' and gives a trivial equal mass decomposition.

We are then able to express the mapping $\Pi$ of a point $\mathbf{q}$ and its environment between Lagrangian space and phase space through mappings of the unit cube centred on $\mathbf{q}$ as $K_3\to\Pi(K_3)\subset\mathbb{R}^6$, i.e. specifically
\begin{equation}
\Pi:\,\mathbf{q}\mapsto \left(\pi_{x_0}(\mathbf{q}),\pi_{x_1}(\mathbf{q}),\pi_{x_2}(\mathbf{q}),\pi_{v_0}(\mathbf{q}),\pi_{v_1}(\mathbf{q}),\pi_{v_2}(\mathbf{q})\right)
\label{eq:map_cube}
\end{equation}
with all $\pi \in P_k$ and $\mathbf{q}\in K_3$. 

Knowing position and velocity at point $\mathbf{q}$ alone without any neighbour information in the unit cube constrains only the 0-order polynomial ($k=0$). This is comparable to the $N$-body case, where also the continuous structure is not retained. In the TetPM method, four points are used. A tetrahedron represents thus the map $\mathbf{q}\mapsto A\mathbf{q} + \mathbf{b}$, where $A$ is a rank 2 tensor and $\mathbf{b}$ a vector, i.e. it describes a simple linear transformation between three dimensional Lagrangian and six-dimensional phase space. The case of $k=1$ corresponds to a trilinear element (made up from 8 points). In TetPM however, the unit cube was decomposed into six tetrahedra, so that one achieves six uni-linear elements per unit cube instead of one tri-linear element as would be the case at $k=1$ order in the approach discussed here. In this paper, we will mainly focus on the $k=2$ tri-quadratic case, which is fully determined by knowing position and velocity at $3^3=27$ locations in the unit cube. {\em Thus, for a system with a given number of degrees of freedom, fewer elements are needed when the order increases, since higher order elements have more internal degrees of freedom.}

Finally, without loss of generality, the approach discussed above for a single unit cube $K_3$ can be extended to all of $\mathbb{R}^3$ by tiling $\mathbb{R}^3$ with such cubes and performing the mapping between Lagrangian space and phase space separately for every tile of Lagrangian space. For cosmological simulations with periodic boundary conditions one uses a finite tiling of the 3-torus.

\subsubsection{Densities from Lagrangian elements}
\label{sec:densities}
In general, the uniform density of a cubical element of Lagrangian space is mapped to a non-uniform density curved patch in configuration space.
We note that the exact density of the mapped unit cube (of mass $m$) is obtained directly from the Jacobian of the configuration space part of the coordinate transformation ${\rm J}\equiv\boldsymbol{\nabla}_\mathbf{q}\otimes\mathbf{x_q}$ (i.e. $J_{ij} \equiv \frac{\partial x_i}{\partial q_j}$) as 
\begin{equation}
\rho \equiv m \,\left| \det {\rm J}\right|^{-1} = m / \sqrt{\left| g \right|},
\label{eq:density_def}
\end{equation}
where the last equality just highlights the connection to differential geometry, i.e. the curved Lagrangian elements define a metric space with the metric ${\rm G} = {\rm J}^T{\rm J}$ and $g = \det\,{\rm G}$, so that caustics correspond to $g=0$. It is obvious then that tetrahedra possess a constant metric, while the higher order polynomial elements are of non-constant metric. Equivalently, density is constant for the tetrahedron, and the reciprocal of a polynomial for $k>0$, which means that elements with $k\ge1$ can explicitly track caustics, while $N$-body and TetPM can not (here caustics arise only when a tetrahedron has vanishing volume).

\begin{figure}
\begin{center}
\includegraphics[width=0.48\textwidth]{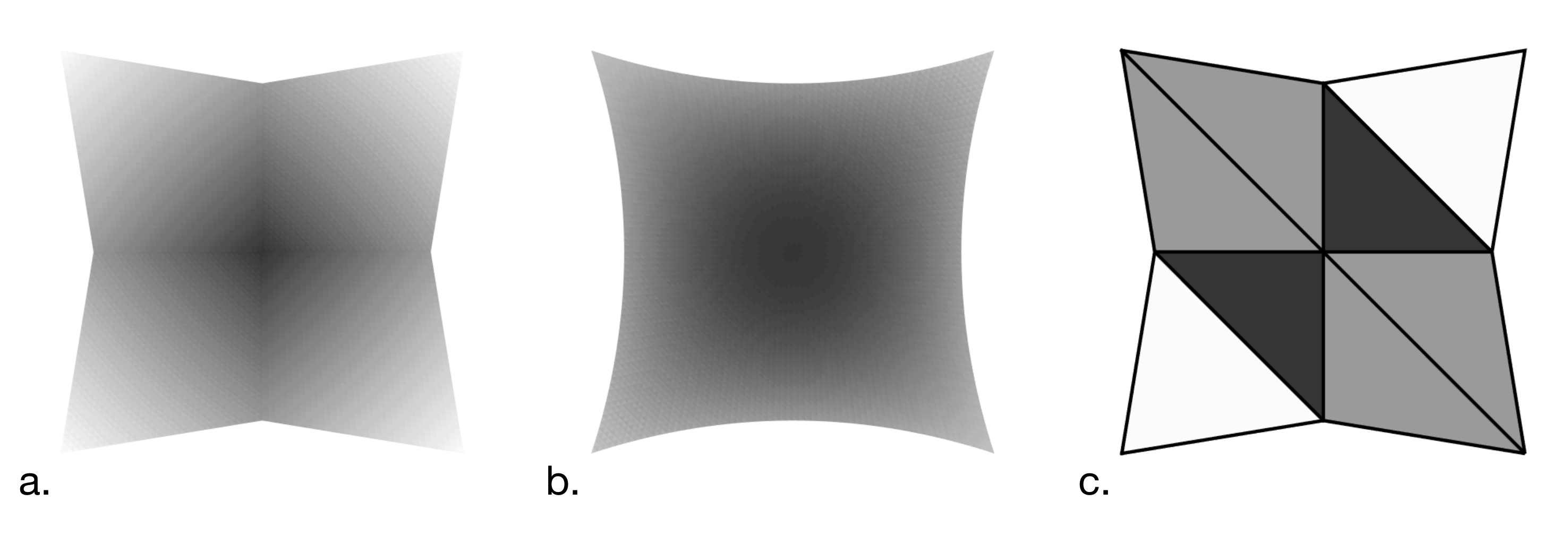}
\end{center}
\caption{
\label{fig:patch_deform}Illustration of the deformation of a Lagrangian unit square represented by $3^2$ supporting points under a contraction inverse proportional to the distance from the centre. The square is mapped using a bi-linear polynomial map (a), a bi-quadratic map (b) and using a simple triangulation (c). 
}
\end{figure}

In Figure~\ref{fig:patch_deform}, we illustrate how a Lagrangian unit square maps to configuration space, and how the inferred density depends on the adopted order of the polynomial. In this two-dimensional example, we represent the Lagrangian unit square by $3\times3=9$ supporting points, and displace the Eulerian coordinates outwards proportional to the distance from the centre of the square. In case a., we show the result of adopting $k=1$ bi-linear polynomials on four sub-squares tiling the full square. In case b., we show the result of adopting $k=2$ bi-quadratic polynomials on the full square. Finally, in case c., we triangulate the full square into eight triangles. We see that the mapped shaped in the two linear methods a and c are identical, but the inferred density is more accurately representing the radial gradient in the bi-linear case than in the tessellation case. The bi-quadratic method leads to parabolic edges and a smooth density distribution in the interior. The resulting density field (before shell-crossing) is piecewise linear in one direction for the tri-linear map, but generally not continuous. The tri-quadratic map, remedies this by allowing for gradients to change smoothly inside of an element. It is always continuous but generally not continuously differentiable. Finally, triangular elements are piecewise constant which leads to a density field where many elements are needed to faithfully describe gradients. Given that all three examples use identical data, the advantage of higher order methods is clearly obvious in this case.  

\begin{figure}
\begin{center}
\includegraphics[width=0.35\textwidth]{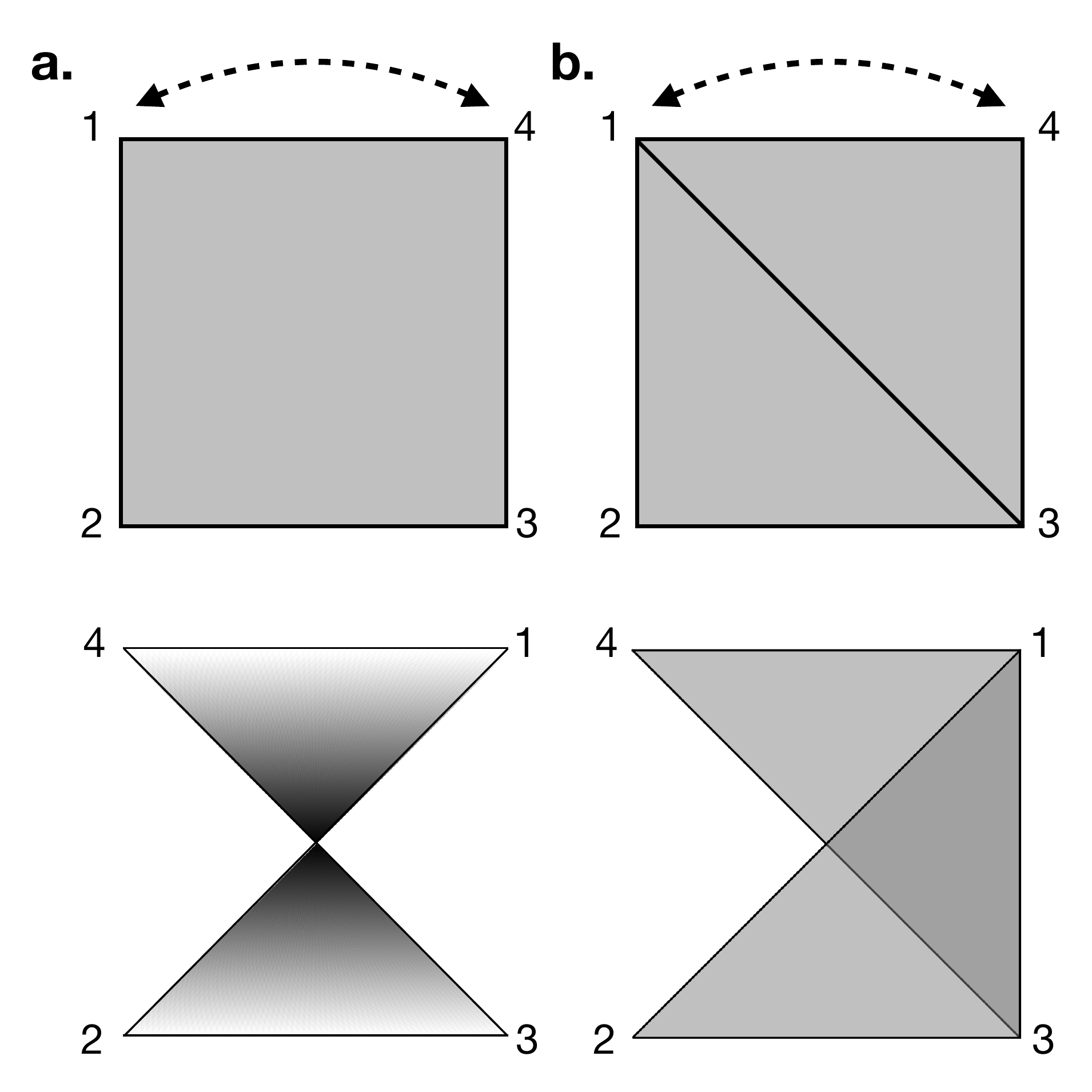}
\end{center}
\caption{
\label{fig:patch_vs_tet}Illustration of the difference between the density induced from the metric of a bilinear form (a), and that of a simple triangulation (b) of the same quad (1234) under a transformation in which the top two vertices (1 and 4) are flipped. The bilinear form leads to a non-convex projection with divergent density at the caustic point, while, naturally, the triangles remain convex, but have always constant density and become overlapping.
}
\end{figure}

We illustrate how polynomial elements trivially capture caustics in Fig.~\ref{fig:patch_vs_tet}, where, by a simple linear map to Eulerian space the top vertices of a square are flipped. In the left panel, a, the result for a tri-linear map is shown. Here, the mapped square has the correct topology and an actual singularity at the centre. This is quite in contrast to what can be achieved with a triangulation, which is shown in the right panels, b. Here, a choice must be made as to how to triangulate the square (in 2D there are two possibilities). The linear map flipping the top vertices then maps each triangle again to a triangle of the same size, which leads to a final density field that has no singularity, but also the mapped square does not have the correct topology (nothing should have been mapped to the region of the dark grey triangle in the bottom right panel). Obviously, for N-body particles, the flipped and un-flipped state are indistinguishable. This anisotropy of a simple triangulation exists of course also in three dimensions. In order to avoid the intrinsic anisotropy inherent in the Delaunay triangulation of the unit cube, HAK13 have proposed an alternative decomposition of the cube into eight overlapping tetrahedra instead of six (the decomposition they employed for the TCM scheme as opposed to the Delaunay that was used for the T4PM; Figure~3 in HAK13, where the TCM decomposition can be thought to be rearranged into one cube). {\em In this paper, whenever we use tetrahedra, we thus employ the same non-tessellating double-covering of the unit cube as in HAK13 to avoid an anisotropic decomposition with tetrahedra.}


\subsection{Time evolution, discretisation errors and refinement strategies}
\label{sec:errors_refinement}
Next, we will consider the time evolution of the Lagrangian elements. We will demonstrate that the evolution equations for the tri-polynomial elements become separable evolution equations for the coefficients that can be solved by simply evolving the supporting points as freely falling flow tracer particles. The representation by finite order polynomials, however, introduces a truncation error that has to be bounded in order to maintain the Hamiltonian character of the system. This then leads to natural refinement criteria that attempt to keep the truncation error within reasonable limits and are necessary to guarantee e.g. energy conservation in general.

\subsubsection{Time evolution of elements} The phase space density is, as in the $N$-body case, conserved along the characteristics given by the canonical equations of motion (cf. eq.~\ref{eq:equations_of_motion}) with the only difference that we are no longer concerned with a discrete set of characteristics but that they have an underlying  manifold structure and thus generalise to
\begin{equation}
\dot{\mathbf{x}}_\mathbf{q} = \mathbf{v}_\mathbf{q}, \quad\textrm{and}\quad
\dot{\mathbf{v}}_\mathbf{q} = -\left.\boldsymbol{\nabla}_x \phi\right|_{\mathbf{x}_\mathbf{q}},\quad\textrm{with }\mathbf{q}\in\mathcal{Q}
\end{equation}
along which $\partial f / \partial t = 0$. 

The time evolution of the phase space elements can be obtained from the time derivative of eq.~(\ref{eq:map_cube}), which contains six time derivatives of polynomials $\pi(\mathbf{q})$. Since the Lagrangian coordinates $\mathbf{q}$ do not depend on time, we are left with time derivatives of the coefficients $a_{\alpha\beta\gamma}$. Furthermore, consistency between the supporting points and the coefficients then requires that the coefficients themselves follow canonical equations, i.e.
\begin{equation}
\dot{\mathbf{x}}_{\alpha\beta\gamma} = {\mathbf{v}}_{\alpha\beta\gamma},\quad \dot{\mathbf{v}}_{\alpha\beta\gamma} = -J^{-1} \mathbf{f}_{\alpha\beta\gamma},
\end{equation}
where $J_{ij}\equiv\partial x_i/\partial q_j$ is the Jacobian and $\mathbf{f} = \boldsymbol{\nabla}_\mathbf{q} \phi$ is the force mapped to Lagrangian space.
It is completely equivalent to evolve the coefficients using their own canonical equations, truncated at some order, or using a number of
freely falling test particles (the ``flow tracers'', their number per element matching the degrees of freedom of the element) 
from which these coefficients can be calculated. This second possibility allows us to use
a set of particles that are evolved like massless $N$-body particles to describe the evolution of each element.

\subsubsection{Discretisation errors}
The time evolution above is exact if the series expansion is not truncated (i.e. for $k\to\infty$). When approximating the evolution equations above by piecewise polynomial expansions, a truncation error is however introduced since the force field across an element is only considered to the order of the polynomial expansion. The force across the element is in general however given as
\begin{equation}
\mathbf{F}_\mathbf{q} = \sum_{\alpha,\beta,\gamma=0}^\infty \mathbf{f}_{\alpha\beta\gamma}\,q_0^\alpha q_1^\beta q_2^\gamma,
\end{equation}
so that the elements capture correctly the evolution of the first $(k+1)^3$ terms, and the error in the momentum update is given by
\begin{equation}
\Delta\dot{\mathbf{v}} = -J^{-1} \sum_{\alpha,\beta,\gamma=k+1}^\infty \mathbf{f}_{\alpha\beta\gamma}\,q_0^\alpha q_1^\beta q_2^\gamma,
\end{equation}
which implies that second order terms are sourced by a non-constant tidal field across the element, third order terms by a non-linearly changing tidal field and so on. This implies that momentum is only conserved at the level of this truncation error. The error is expected to be small when the potential is smooth across elements, which is usually true when the elements are small, but is certainly an invalid assumption at later times. Estimating the magnitude of terms of order $k+1$ and larger across elements thus naturally leads to a refinement criterion, so that when an element is split into smaller elements, the truncation error remains bounded. We will elaborate on this possibility next.

\noindent{\underline{\em Refining on force:\,}\,\,} As we have just discussed, errors arise when the variation of the gravitational force across the element is not accurately captured. To turn this error source into a criterion for adaptive refinement, we calculate the force at using two different orders of the interpolating polynomial, i.e. we calculate
\begin{equation}
\mathbf{F}_\mathbf{q} = \sum_{\alpha,\beta,\gamma=0}^k \mathbf{f}_{\alpha\beta\gamma}\,q_0^\alpha q_1^\beta q_2^\gamma\quad\textrm{and}\quad\mathbf{F}'_\mathbf{q} = \sum_{\alpha,\beta,\gamma=0}^{k'} \mathbf{f}_{\alpha\beta\gamma}\,q_0^\alpha q_1^\beta q_2^\gamma,
\end{equation} 
with $k'>k$, then we can estimate the maximum relative force error and attempt to keep it bounded
\begin{equation}
\Delta f \simeq \max_q \left\| \mathbf{F}_\mathbf{q}-\mathbf{F}'_\mathbf{q}\right\|/\left\|\mathbf{F}_\mathbf{q}\right\| < \Delta f_{\rm max} / 2^\ell,
\end{equation}
where $\ell=0,1,\dots$ is the refinement level of the element. Specifically, in our implementation, we choose $k'=2k$ and chose not to determine the maximum but approximate the problem by evaluating $\Delta f$ at half-points, i.e. at the locations where new supporting points would be placed if the element were to be refined (cf. Fig.~\ref{fig:coarse-fine}). E.g., for the tri-quadratic element, we compute the tri-quartic interpolant by simply combining 8 tri-quartic elements together into a fundamental grid unit of $5\times5\times5$ supporting points which provides us with the necessary amount of points to compute the tri-quartic polynomial. 

\noindent{\underline{\em Refining on high order derivatives:\,}\,\,} Alternatively, and similar in spirit to the refinement criterion based on second derivatives proposed by \cite{Lohner1987}, we can keep errors arising from derivatives of order $k+1$ across our elements bounded by using a criterion that splits the Lagrangian cubes when the dimensionless ratio of derivatives of order $k+1$ to derivatives of order $k$ exceeds a specified threshold. Since we will focus on the case of $k=2$ in the remainder, we use the ratio between third and second derivatives of the form
\begin{eqnarray}
\frac{\eta}{\delta} & = &  \frac{1}{2^\ell}\frac{ \left| -\frac{1}{2}f_{i-2}+f_{i-1}-f_{i+1}+\frac{1}{2}f_{i+2}  \right| }{ \left| -\frac{1}{2}f_{i-2}+f_{i-1} \right| + \left|-f_{i+1}+\frac{1}{2}f_{i+2}\right|  + \epsilon F}\\[8pt]
 && F = \left|\frac{1}{2}f_{i-2}\right| +  \left| f_{i-1} \right| + \left| f_{i+1}\right|+ \left|\frac{1}{2}f_{i+2}\right|, \nonumber
\end{eqnarray}
where indices $i$ run over the supporting points in Lagrangian space. Note that in three dimensions, we calculate the derivatives $\eta_i$ and $\delta_i$ along dimension $i$ and then evaluate the quantity
\begin{equation}
\Theta = \left( \frac{\eta_0^2+\eta_1^2+\eta_2^2}{\delta_0^2+\delta_1^2+\delta_2^2}\right)^{1/2}
\label{eq:refine_derivative}
\end{equation} 
A Lagrangian element is selected for refinement into smaller elements when $\Theta>\Theta_{\rm max}$. We typically use $\epsilon=0.01$ and thresholds are $0<\Theta_{\rm max}<1$,  where low values lead to more aggressive refinement, but typically $\Theta_{\rm max}\simeq 0.1$.

Other refinement criteria are of course possible and might possibly perform better than what we consider in this paper. What is the optimal criterion, i.e. minimises errors with the smallest number of flagged elements for a given problem, is an open question. It will thus be interesting to consider alternatives to the two presented above in the future. 


\subsection{Adaptive refinement of elements}
We next discuss the adaptive splitting of Lagrangian elements. We adopt an oct-tree based refinement strategy, meaning that a cubical element can be split into eight smaller cubical elements coextensive with the original element in Lagrangian space. Splitting this way in Lagrangian space automatically implies an equal-mass splitting and exact mass conservation.

\begin{figure}
\begin{center}
\includegraphics[width=0.48\textwidth]{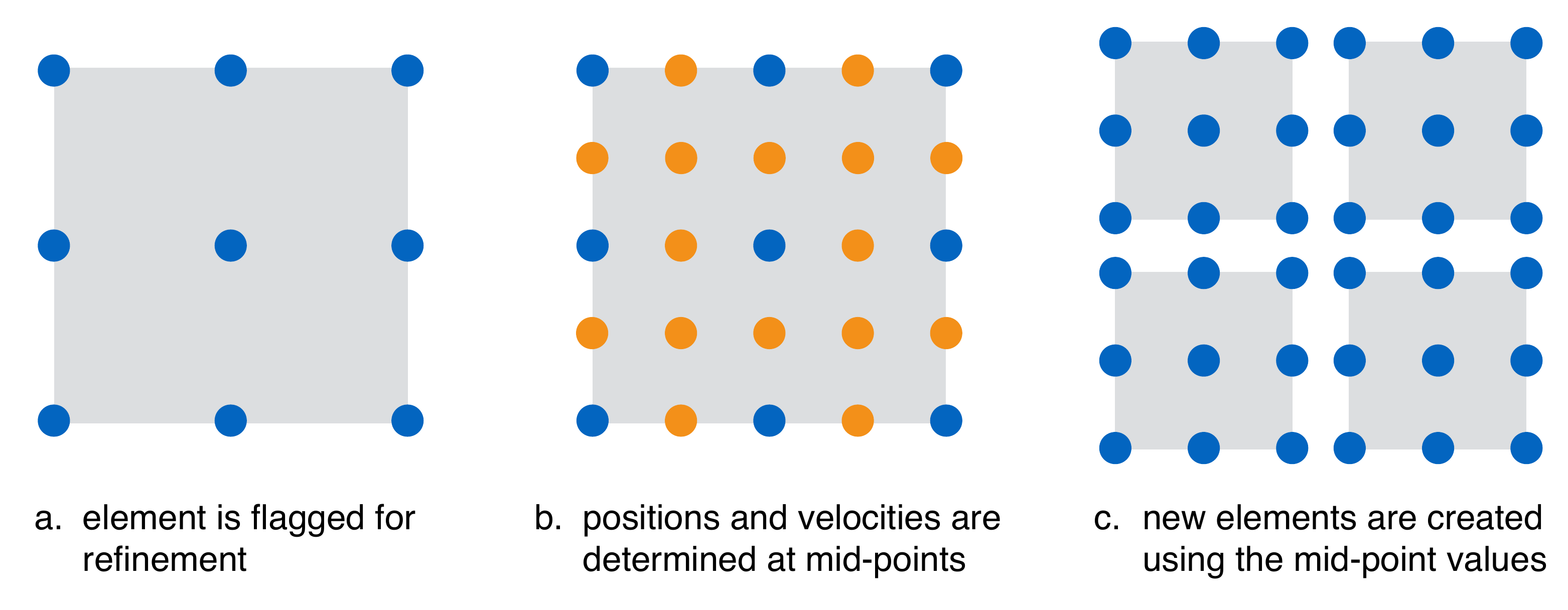}
\end{center}
\caption{
\label{fig:coarse-fine}The three steps followed in our oct-tree based refinement of elements in Lagrangian space: Element flagging by a refinement criterion (left), interpolation to mid-points (middle) and splitting of the element into eight elements of equal mass (right panel).
}
\end{figure}

In order to illustrate the procedure we follow, we discuss the steps involved using the Lagrangian element in two dimensions shown in Figure~\ref{fig:coarse-fine}.  The steps followed during a refinement step are then as follows (and trivially generalise to three dimensions):

\begin{enumerate}[1.]
\item The refinement criterion is evaluated for each element and if it is fulfilled, the element is flagged for subsequent refinement (cf. left panel of Fig.~\ref{fig:coarse-fine}).
\item A regularisation step is performed to ensure that neighbouring elements differ by at most one level of refinement. All elements that need to be refined to maintain regularity are also flagged.
\item For all flagged elements, new supporting points are inserted by evaluating eq.~(\ref{eq:map_cube}) at half-point locations (yellow points in middle panel of Fig.~\ref{fig:coarse-fine}).
\item The element is split into eight elements of $1/8$th the mass of the original element using the interpolated midpoints (cf. right panel of Fig.~\ref{fig:coarse-fine}).
\end{enumerate}

The algorithm outlined above is guaranteed to provide a covering of Lagrangian space at all times (i.e. no holes will appear and mass is explicitly conserved), and changes in refinement level are constrained to be at most one between neighbouring cells. {\em This procedure thus implements a tree-based AMR method in Lagrangian space.} Since this AMR structure can be represented by an oct-tree, all operations on the tree can be encoded using integer arithmetic (we represent Lagrangian space coordinates by three 20bit integer numbers that can be stored in one 64bit integer) allowing for fast identification of neighbours without explicitly storing pointers.

All flow tracers are allowed to move freely, except those that occupy the faces between elements of different resolution. {\em At these coarse-fine boundaries, flow tracers are constrained to remain in Lagrangian space at the mid points between coarser particles.} This can be achieved by using an acceleration interpolated from the flow tracers shared between the coarse and the fine element to the half points instead of evaluating their actual acceleration from the gravitational force at their location. This ensures that also no holes appear when the Lagrangian manifold is mapped to Eulerian space.


\subsection{Gravitational Evolution}
We now describe how we compute the gravitational force field sourced
by the Lagrangian elements and discuss how to follow their time evolution.
We will discuss how we compute the total density field $\varrho(\mathbf{x}) = \sum_\epsilon \rho_\epsilon(\mathbf{x})$,
where $\epsilon$ indexes all Lagrangian elements. Having an estimate of
the total density field, it is then possible to solve Poisson's equation
$\boldsymbol{\nabla}_\mathbf{x}^2\phi = 4\pi Ga^{-1}\left(\varrho - \bar{\varrho}\right)$
and compute the gravitational accelerations.

\subsubsection{Approximating the projected mass distribution of Lagrangian phase space elements}

\begin{figure}
\begin{center}
\includegraphics[width=0.48\textwidth]{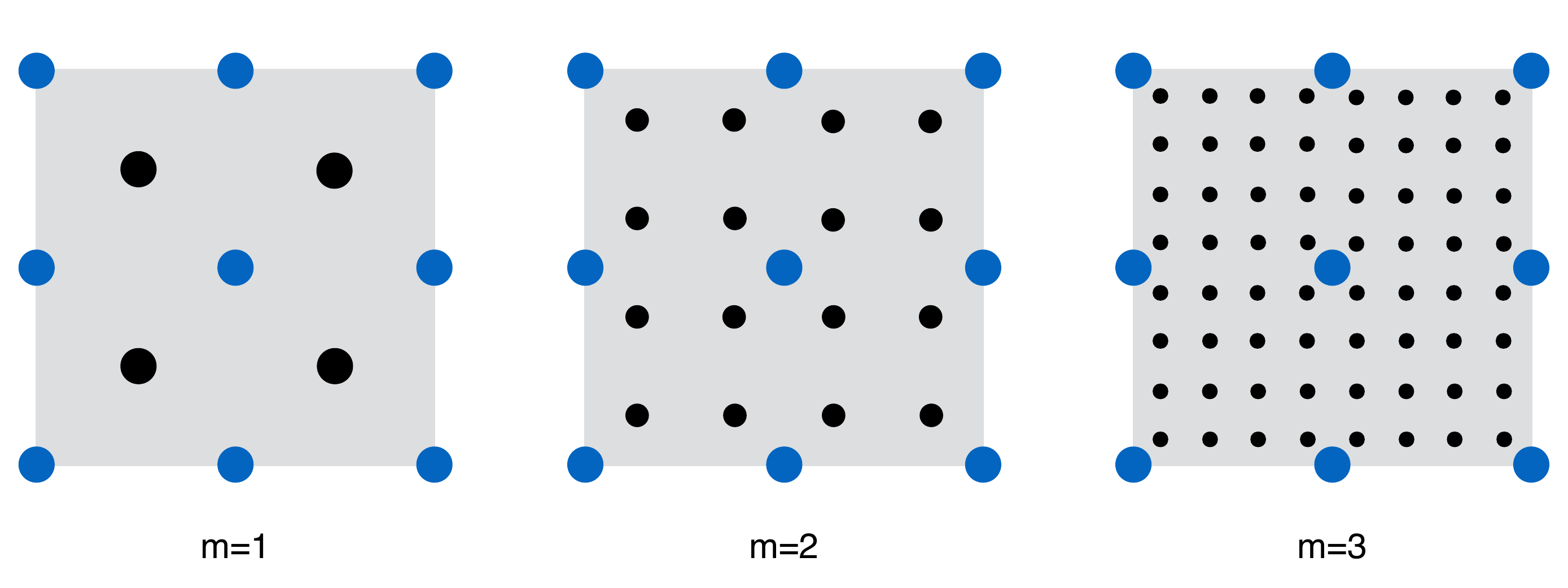}
\end{center}
\caption{
\label{fig:mass-deposit}Mass deposition through oct-tree based splitting in
Lagrangian space. The mass distribution of elements is approximated by
recursively split mass carriers (black dots at one, two and three levels of
recursion from left to right). Their position is determined using the local
mapping between Lagrangian and Eulerian space which is determined from the
supporting points (``flow tracers'', in blue).  }
\end{figure}

A basic element consists, in general, of a curved hexahedron whose inner density
field is described by a rational function. In principle it may be possible to directly
compute in some simple cases the force exerted by these geometrical objects. In practice, however,
this would be computationally extremely expensive (if possible at all). A more practical alternative is
to project the Lagrangian elements onto a regular mesh in Eulerian configuration space and compute the forces by solving
Poisson's equation using Fourier or relaxation methods.

Choosing such an approach, the elements need to be deposited onto a grid. In general there is
no known method to integrate an arbitrary rational function over the intersection of
curved hexahedra and cubes (i.e. grid cells). An alternative is to approximate
them by polyhedra and then decompose them into a set of tetrahedra, for which an exact deposit is possible
\citep{Powell:2014}. Another alternative, which we adopt here,
is to recursively decompose, in Lagrangian space, the mass of the hexahedra
into equal-mass particles located at the centre of mass of each subvolume. This set
of mass carriers is then mapped to Eulerian space using eq.~(\ref{eq:map_cube})
and deposited onto the mesh as point masses. We describe this approach in more
detail below.

Our cubical Lagrangian elements are fully described by a set of $3^n$ points in $n$ dimensions,
spanning $2^n$ linear unit volumes, or 1 quadratic unit volume (cf.
Fig~\ref{fig:mass-deposit}, where the blue points indicate the supporting
points, or ``flow tracers'' and the shaded grey area the Lagrangian volume
spanned by them).  In contrast to the tetrahedral elements of TetPM (which have
constant density), our tri-linear and tri-cubic (and all higher order) mapping
of these elements results in a non-uniform density. Fortunately, their
projection can be trivially approximated with a simple oct-tree based particle
deposit scheme. Since the density in Lagrangian space is uniform, we perform an
oct-tree subdivision of the Lagrangian unit cube $K_3$ up to some level $m$
that decomposes each element into $2^{3m}$ discrete particles (or ``mass
carriers'') as illustrated in Fig.~\ref{fig:mass-deposit}.  For every element,
we have the $(k+1)^3$ supporting points $\Sigma_k$, cf.
eq.~(\ref{eq:supporting_points}), that fully determine the mapping $\Pi$,
eq.~(\ref{eq:map_cube}). We then represent the element of total mass $m$ at
approximation level $m$ through a space partitioning into octs with mass
$m/2^{3m}$ and Lagrangian centres of mass at the locations $\Xi_m\subset K_3$

\begin{equation}
\Xi_m \equiv \left\{ \left. \left(\frac{2 i_0-1}{2^{m+1}},\frac{2 i_1-1}{2^{m+1}},\frac{2 i_2-1}{2^{m+1}}\right)\,\,\right|\,\, i_j=1,\dots,2^m\right\}.
\end{equation}

The mapping to Eulerian space for all points in $\Xi_m$ is then given by simply
applying the mapping $\Pi$, i.e. $\Pi(\Xi_m)$, that has been determined using
the supporting points $\Sigma_k$. In practice, since the points at which $\Pi$
needs to be evaluated are given by the simple oct-tree structure, the
coefficients of the 3-variate polynomials can be pre-calculated. 

One apparent disadvantage of our approach is that mapped particles have a
deformed grid regularity. For completeness, we mention three alternatives to
deposit particles at the end of the recursive decomposition. 

\begin{enumerate}[i.]
\item A very smooth deposition could be achieved by triangulating the final
oct  and depositing the resulting tetrahedra using an exact deposition algorithm
\citep[cf.][]{Powell:2014}.

\item A glass particle distribution  \citep[cf.][]{White1996} can be used to map 
the elements to Eulerian space suppressing possible
Moir\'e between the density mesh and the deposited particles (i.e. $\Xi_m$ is
replaced with a glass distribution in $\left[0,1\right]$). 

\item A random sampling is also possible, where $\Xi_m$ is replaced with an
arbitrary number of uniformly distributed random points drawn from
$\left[0,1\right]$. This approach is not suitable to sample the mass
distribution in time evolving simulations since a single random realisation is
not force free (in contrast to the glass and the hierarchical lattice).
However, this approach is favourable for the creation of images since Poisson
sampling gives the visual impression of a smoother field than the other
approaches using the same number of sampling points.
\end{enumerate}

\noindent All of these three alternatives result into no or fewer symmetries, suppressing
possible Moir\'e between the density mesh and the deposited particles. Currently, we have
successfully implemented ii) and iii) in our code. However, we have tested that
this does not affect our results in practice, thus we will not employ the
simple grid deformation in the reminder of the paper.

\subsubsection{Force Calculation and Time Stepping}

The particles in $\Xi_m$ can deposited using any of the known
particle-mesh charge assignment schemes \citep[cf.][]{Hockney1981}. We have
adopted the cloud-in-cell (CIC) deposit in this paper for the pseudo-particles
$\Xi_m$ determined as described above. Once the density field is obtained, we solve Poisson's equation using
a standard Fourier method. We first perform a Fast Fourier Transform of the field, then compute the gravitational
potential by multiplying with Green's function for a fourth order finite-difference Laplacian, i.e.

\begin{equation}
\begin{split}
G(k) =  & \left[\,  \cos(2\,k_x) - 16 \cos(k_x) + 15 \right.\\ 
                   + &  \left.    \cos(2\,k_y) - 16 \cos(k_y) + 15\,\right.  \\ 
                   + &   \left.  \cos(2\,k_z) - 16 \cos(k_z) + 15\,\right]\,/\,6.
\end{split}
\end{equation}

\noindent We then perform the backward Fourier transform and apply a 4th order
finite difference gradient to compute the gravitational force field. We finally
interpolate this field to the position of the supporting points and advance
their positions and velocities just as would be done for standard $N$-body particles.

We note that our implementation of the force algorithms do not require all mass
carriers to be concurrent in memory. This is a great advantage in keeping 
the memory footprint of our code low since the ratio of flow to mass tracers will
be small.

Once forces are computed, we adopt a leapfrog kick-drift-kick time integration
scheme. Since our forces are given by a PM algorithm, we employ a global 
timestep set by the minimum of the timestep assigned to flow tracers, which
in turn is set by a Courant criterion:

\begin{equation} 
t = min\left[C\,\frac{\Delta x}{u_i} \right]
\end{equation} 

\noindent where $\Delta x$ is the FFT cell size, $u_i$ is the magnitude of the
velocity for the $i$-th particle, and $C$ is a parameter which we usually set
to somewhat less than unity.

For completeness, we mention that we have also implemented an oct-tree
algorithm, which allows us to have a very high force resolution and also the
implementation of individual time steps for every flow tracer. The idea is to
build a full oct-tree down to some specified node length, and then use flow
tracers to refine the tree topology in a given simulation. Then, the moments of
the multipole expansion of the tree force of each node are computed using the
mass field given by all mass carriers. In the remainder of this paper
we, however, only employ forces computed by the PM algorithm to avoid possible
artefacts due to spatially-correlated force errors introduced by the use
of finite terms in the multipole expansion of the force field.


\subsection{Implementation and Parallelisation Strategies}

Interesting problems in cosmological structure formation require a full
exploitation of state-of-the art supercomputer centres. This in turn requires
algorithms capable of being efficiently executed on up to thousands of processor
cores with a mixture of shared and distributed memory, and for billions of
resolution elements. Thus, we have put particular emphasis on these aspects in
our code.

We have implemented our algorithms in the distributed-memory parallel code
{\tt L-Gadget3}. The {\tt L-Gadget3} code is a memory efficient version of
{\tt P-Gadget3} \citep{Springel2005b,Springel2008} specially tailored for dark
matter only simulations and that was successfully executed on 12,000 computer
cores for the MXXL project \citep{Angulo2012a}. Most of the parallel algorithms
can be directly used for our Lagrangian element method, but there are two
aspects noteworthy.

The first is the one-time construction of Lagrangian patches from a set of simulation
particles. For this, we start by performing a decomposition of the computational
domain using Lagrangian (as opposed to Eulerian) coordinates. This makes most 
Lagrangian patches to have locally all their supporting points. For those lying 
in the surface of the domain boundary this is not true and they will need to 
communicate with other processors and import the required set of vertex. This 
can be done in a very efficient manner by exploiting the fact that there is a 
unique relation between a particle ID, its Lagrangian patch, and the MPI task 
hosting it.

The second aspect is that our fundamental data structure are Lagrangian
patches, on which an Eulerian domain decomposition is performed. We keep a
locally complete set of supporting points for these structures, which implies
a slight overhead in terms of memory (since the boundary supporting points are
duplicated) and slight suboptimal work/load balance. However, our choice has
the huge advantage of allowing all quantities related to Lagrangian patches to
be computed locally, without requiring any inter-node communication. This
makes the creation of mass carriers embarrassingly parallel, while their
subsequent deposit to the mesh requires communication.





\section{Testing Adaptive Refinement: Phase Mixing in a Fixed Potential}
\label{sec:testing}
In a first test problem, we evolve a cube of cold collisionless material without self-gravity in a fixed potential. This will allow us to study the general feasibility of adaptive refinement as well as the impact of the refinement criterion on the accuracy of the subsequent evolution. After this, in the subsequent sections, we then turn to test problems with self-gravity in one and two dimensions.

As discussed in HAK13, the most important shortcoming of TetPM is the inability to follow the dark matter sheet when strong mixing occurs (e.g. in the inner parts of haloes). In this case, the surface of the sheet in phase space grows rapidly (see also below) by being continuously stretched, and a given small number of points is insufficient to describe its evolution over longer times. New points need to be inserted in order to follow the motion of the stretched regions accurately.

To provide a robust and simplified framework to test the performance in mixing situations, we study a cube $K$ of collisionless, cold but not self-gravitating material orbiting in a non-harmonic potential. The cube is initially represented by $5^3$ particles spanning $8\times 4^3$ tetrahedral, $4^3$ tri-linear or 8 tri-quadratic elements respectively. For comparison, we will compare to the high-resolution $N$-body solution in which we sample the cube by $128^3$ particles. The cube is given an initial uniform velocity that is offset from the centre of the potential (i.e. every point in it has non-zero angular momentum). Specifically, we consider a constant Plummer gravitational potential for which the acceleration field is given by
\begin{equation}
\ddot{\mathbf{x}}_i = -\frac{\alpha\,\mathbf{x}_i }{\left(\left\| \mathbf{x}_i \right\|^2 + \epsilon^2\right)^{3/2}}.
\label{eq:plummer_pot}
\end{equation}
with some $\alpha$ and $\epsilon$ whose specific values are irrelevant in our case as we only want to study the qualitative behaviour. Since the potential is constant in time, the motion of any point $\mathbf{q}\in K$ in phase space is  determined uniquely by its initial coordinates in phase space $(\mathbf{x}_\mathbf{q}(0),\mathbf{v}_\mathbf{q}(0))$. Hence, the choice of the Lagrangian elements used to represent the orbiting cube has no impact on the dynamics of the vertices. The order of the elements will however impact the distribution of matter between the vertices, as well as where new vertices are inserted if adaptive refinement is used. This simple test problem will thus allow us to study a few key diagnostics: (1) how the order of the Lagrangian elements impacts where mass is deposited over time and how well energy is conserved for the elements, (2) which refinement criteria allow to follow the mixing well and allow an energy conserving scheme, as well as (3) which refinement criteria yield a better tracking of the evolving phase space sheet with the smallest number of newly inserted vertices.

\subsection{Phase mixing without refinement}
\label{sec:mixing_noref}
\begin{figure} 
\begin{center}
\includegraphics[width=0.4\textwidth]{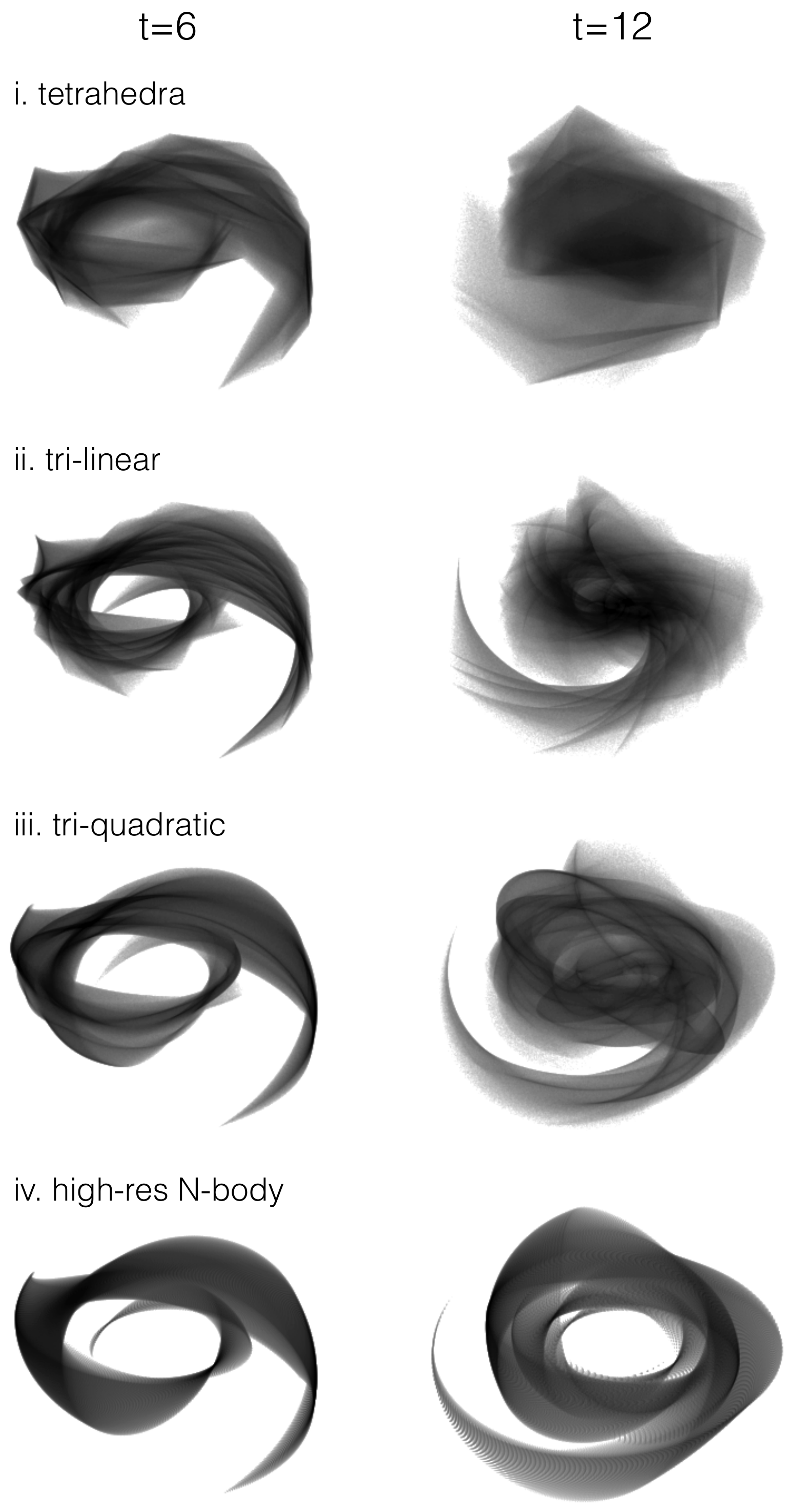}
\end{center}
\caption{
\label{fig:fixed_potential_ref_images} Images of a uniform cube phase-mixing in a Plummer potential at $t=6$ and at $t=12$. The top three panels use the exact same $125$ supporting points but differ in the order of the Lagrangian elements determined by them (i.e. they have the {\em same} degrees of freedom, but differ in the number of elements). The bottom row shows the result of sampling the cube with more than 2 million N-body particles. All Lagrangian element schemes diffuse mass towards the center, but the loss of energy and resulting 'overmixing' is smallest for the tri-quadratic elements (as quantified in Fig.~\ref{fig:fixed_potential_noref}).
 }
\end{figure}

\begin{figure} 
\begin{center}
\includegraphics[width=0.4\textwidth]{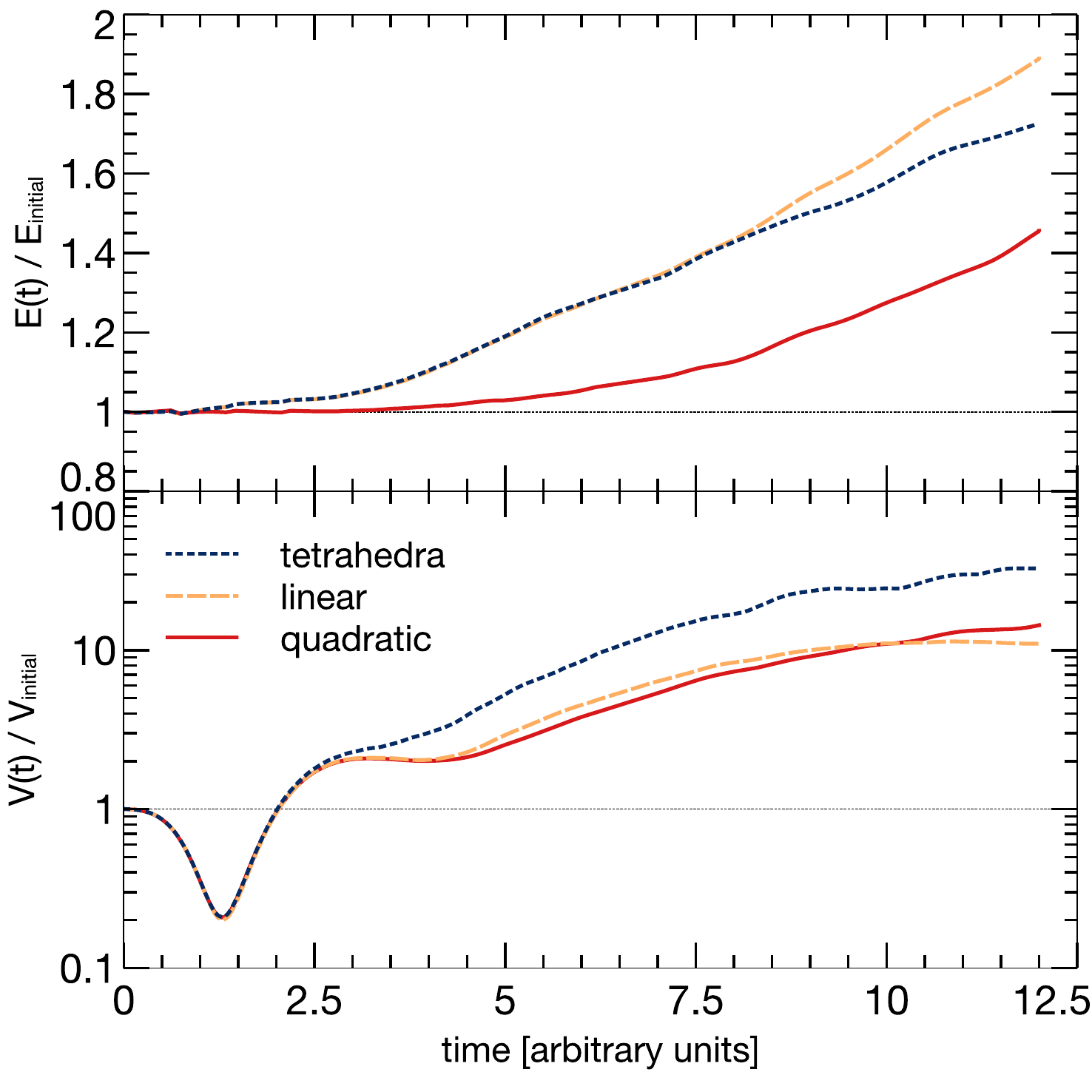}
\end{center}
\caption{
\label{fig:fixed_potential_noref} Evolution of the total energy (top) and volume (bottom panel) of a uniform density cube orbiting in a static Plummer potential, leading to phase mixing. The elements are {\em not} adaptively refined in this test, i.e. the degrees of freedom in the system are kept constant. In both panels we show results for tetrahedral (dotted blue), tri-linear (dashed yellow) and tri-quadratic (solid red) Lagrangian elements. The significantly higher volume for the tetrahedra is spurious and should be divided by $\sim2$ at late times (see text for details).
 }
\end{figure}

We first demonstrate once again that in mixing situations the Lagrangian element schemes lead to mass deposit towards the centre of the potential, leading to a violation of energy conservation. In Figure~\ref{fig:fixed_potential_noref}, we show the sheared cube at two times $t=6$ and $t=12$ (left and right columns, respectively) for all schemes: tetrahedral, tri-linear, tri-quadratic Lagrangian elements based on only $5^3$ supporting points to describe the initial cube as well as the respective $N$-body solution using $128^3$ particles to sample the initial cube (from top to bottom). We note that generally higher order schemes, as expected, help to represent the true mass distribution more faithfully, with the tri-quadratic scheme providing a reasonably good description at $t=6$, where no mass is deposited into disallowed orbits close to the centre. This is not the case for the linear schemes where significant diffusion towards the centre has occurred already by $t=6$. At $t=12$, the situation is even more dramatic, as all schemes now deposit mass into the very centre of the potential, with the tetrahedra even assigning most mass to the central region. This is exactly the situation that lead to the biased central densities of haloes discussed by HAK13. While the curved elements help to some degree to faithfully describe the wrapping of the sheet, it is obvious that a refinement technique is needed to fully resolve diffusion towards the centres of potentials and thus maintain energy conservation. 

To quantify the visual impression gained from Fig.~\ref{fig:fixed_potential_noref}, we calculate the total energy of the cube over time
\begin{equation}
E_{K} = \bar{\rho} \int_K \left[  \frac{v_\mathbf{q}^2}{2} + \phi(\mathbf{x}_\mathbf{q}) \right] {\rm d}^3 q,
\end{equation}
which can be particularly easily evaluated for an analytic potential and is conserved if the potential is static as in our case. Furthermore, the degree of sheet wrapping can be quantified by the total volume of the cube as given by
\begin{equation}
V_{K} = \int_K \sqrt{\left| g \right|}\,{\rm d}^3 q.
\end{equation}
In Figure~\ref{fig:fixed_potential_noref}, we show the evolution of $E$ and $V$ with respect to their initial values. We find that at early times, the tetrahedral and tri-linear elements lose energy at the same rate, for $t\gtrsim 7$, the tetrahedra lose somewhat less than the tri-linear. At all times, the tri-quadratic scheme conserves energy significantly better, but still loses energy. The sheet volume is identical for all methods for $t\lesssim 2.5$ and disagrees after, with tri-linear and tri-quadratic roughly following each other and tetrahedra estimating a significantly larger volume (roughly a factor of two). This is however simply due to the fact that the tetrahedral decomposition is not an actual tessellation to avoid intrinsic anisotropies (cf. the discussion in Section~\ref{sec:densities}). It is obvious from these results, that adaptive refinement of the elements will be required to both conserve energy and arrive at a converged answer for the volume of the sheet.


\subsection{Phase mixing with refinement}
\label{sec:mixing_ref}
\begin{figure} 
\begin{center}
\includegraphics[width=0.4\textwidth]{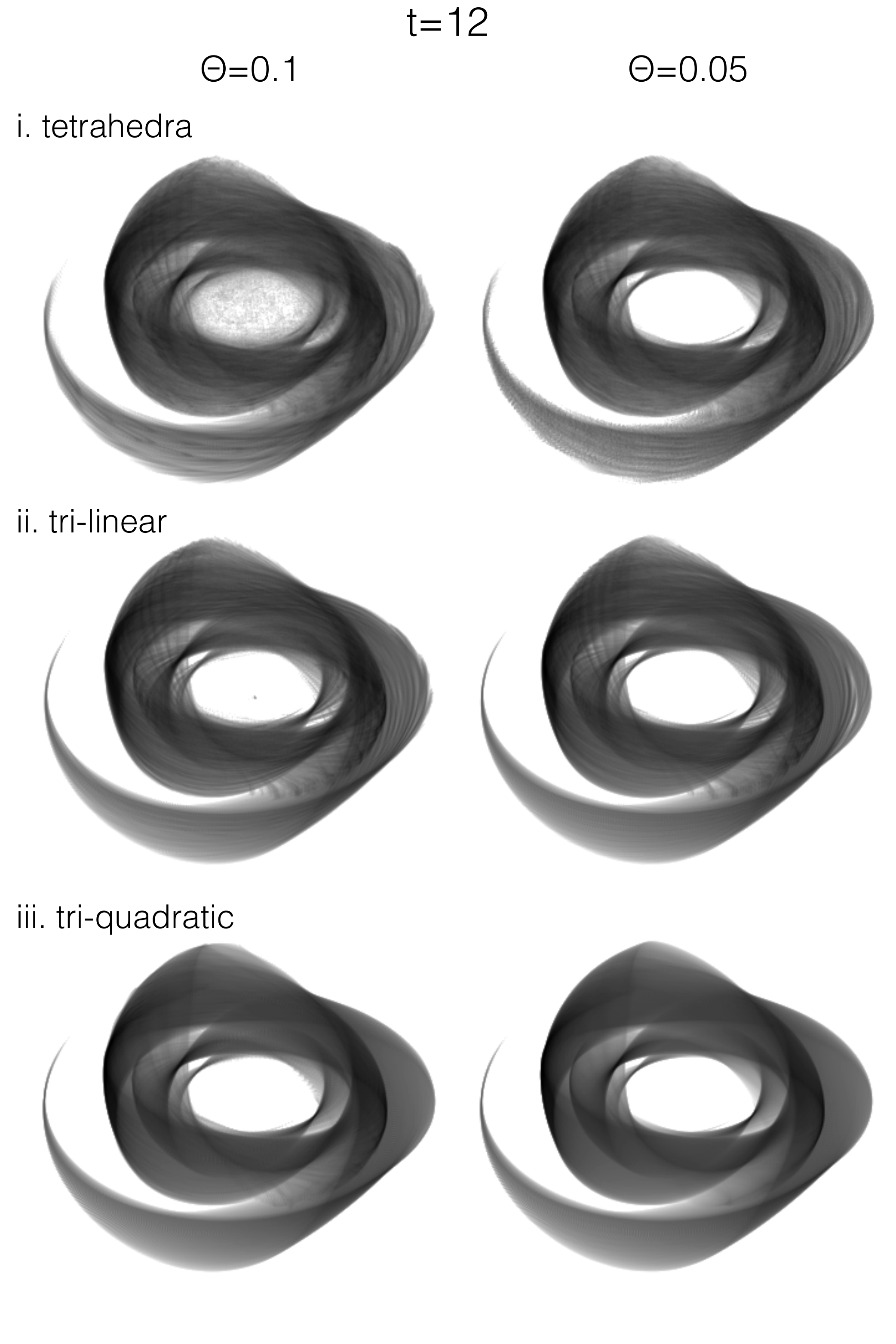}
\end{center}
\caption{
\label{fig:fixed_potential_ref_images1} Images of the mixing cube from Figure~\ref{fig:fixed_potential_ref_images} at $t=12$ with dynamical adaptive refinement using the criterion from eq.~(\ref{eq:refine_derivative}). Dynamical refinement increases the number of degrees of freedom available to the system (beginning from the same 125 as in Fig.~\ref{fig:fixed_potential_noref}). Thanks to the adaptively inserted vertices, all schemes now reproduce the mass distribution at $t=12$ reasonably well, with higher order schemes still performing better. Energy is increasingly better conserved (as quantified in Fig.~\ref{fig:fixed_potential_ref}), for the tri-quadratic elements to better than 0.1 per cent. See text for a detailed discussion.
 }
\end{figure}

\begin{figure} 
\begin{center}
\includegraphics[width=0.4\textwidth]{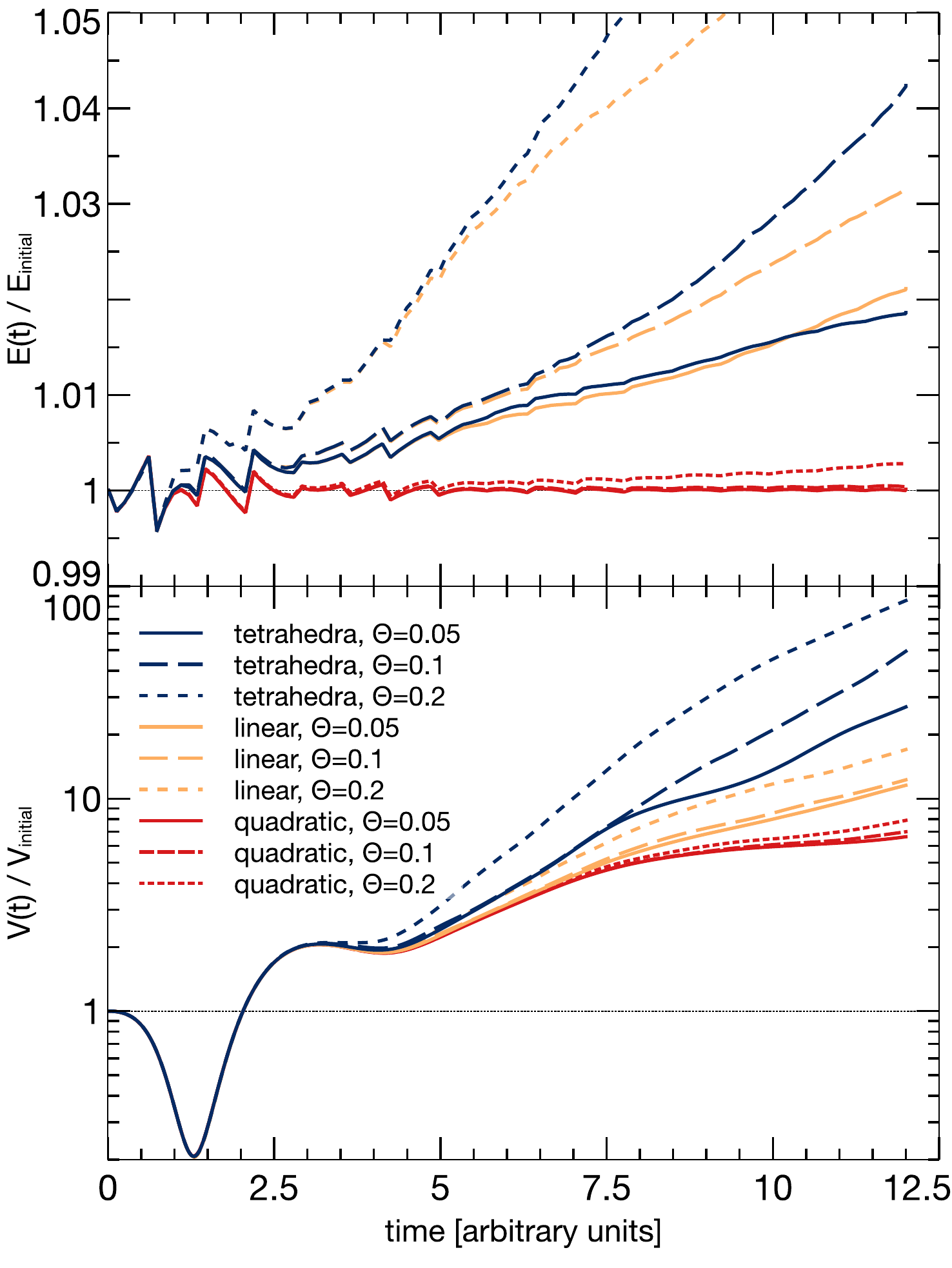}
\end{center}
\caption{
\label{fig:fixed_potential_ref} Evolution of the total energy (top) and volume (bottom panel) of the uniform density cube orbiting in a static Plummer potential {\em with} dynamical adaptive refinement (to be compared with Fig.~\ref{fig:fixed_potential_noref} that shows the same results without refinement). For each of the element types, we show the evolution for various choices of the refinement threshold parameter $\theta$ (c.f. eq.~\ref{eq:refine_derivative}). With the tri-quadratic elements energy is well conserved to a degree controllable by the refinement threshold $\theta$.
 }
\end{figure}

We now allow for adaptive refinement using the refinement criterion based on limiting high-order polynomial contributions to the represented velocity field as expressed in equation~(\ref{eq:refine_derivative}). We have tested also refining on forces, but found that a significantly larger amount of refinement was needed to arrive at a solution of comparable quality\footnote{This can be understood simply from the difference in location w.r.t. the potential centre where the respective criteria trigger refinement. The velocity refinement inserts new vertices preferentially at large radii where large differences in velocity for an element occur first. Quite in contrast to this, the force refinement will split the innermost elements first since they experience stronger force gradients. The latter proved to be less optimal in this case.}.

First, we will now investigate the impact of the order of the element as well as the threshold used in the refinement criterion on the evolution of the cube. In Figure~\ref{fig:fixed_potential_ref_images1}, we show the distribution of mass at time $t=12$ (to be compared to the right column of Fig.~\ref{fig:fixed_potential_ref_images}) for the three Lagrangian element types as well as refinement thresholds of $\theta=0.1$ and $\theta=0.05$. We note that adaptive refinement has significantly improved the agreement with the density distribution of the $N$-body reference solution for all element types with increasing agreement, as expected, with higher order and smaller thresholds $\theta$. In fact, the tri-quadratic solution with threshold 0.05 outshines the reference solution which is grainy due to the $N$-body sampling, while both use about two million points to represent the solution. For $\theta=0.05$ the lower order elements refined to at least this number of vertices, while still suffering from slight density fluctuations. For $\theta=0.1$, all elements refined to make use of a few hundred thousand vertices to represent the solution. We complement this visual inspection with a quantitative study of the total energy and the sheet volume in Figure~\ref{fig:fixed_potential_ref}. In all cases (we also show results for $\theta=0.2$ in this figure), energy conservation is greatly improved, but remains only for the tri-quadratic elements at the sub-percent level by $t=12$. Interestingly, the different elements seem to converge to different asymptotic curves for the evolution of the sheet volume when the threshold $\theta$ is reduced. The sheet volume is a very sensitive diagnostic since all perturbations and numerical errors will enhance the sheet growth rate. The increase in sheet volume is entirely due to the density inhomogeneities visible in Figure~\ref{fig:fixed_potential_ref_images1} for the tetrahedral and linear elements. While it may be possible to find refinement criteria more suitable to the low order elements, we conclude at this stage that the tri-quadratic elements are significantly improving energy conservation and quality of the solution with a significantly smaller number of necessary vertices to be inserted via adaptive refinement.

\begin{figure*} 
\begin{center}
\includegraphics[width=0.9\textwidth]{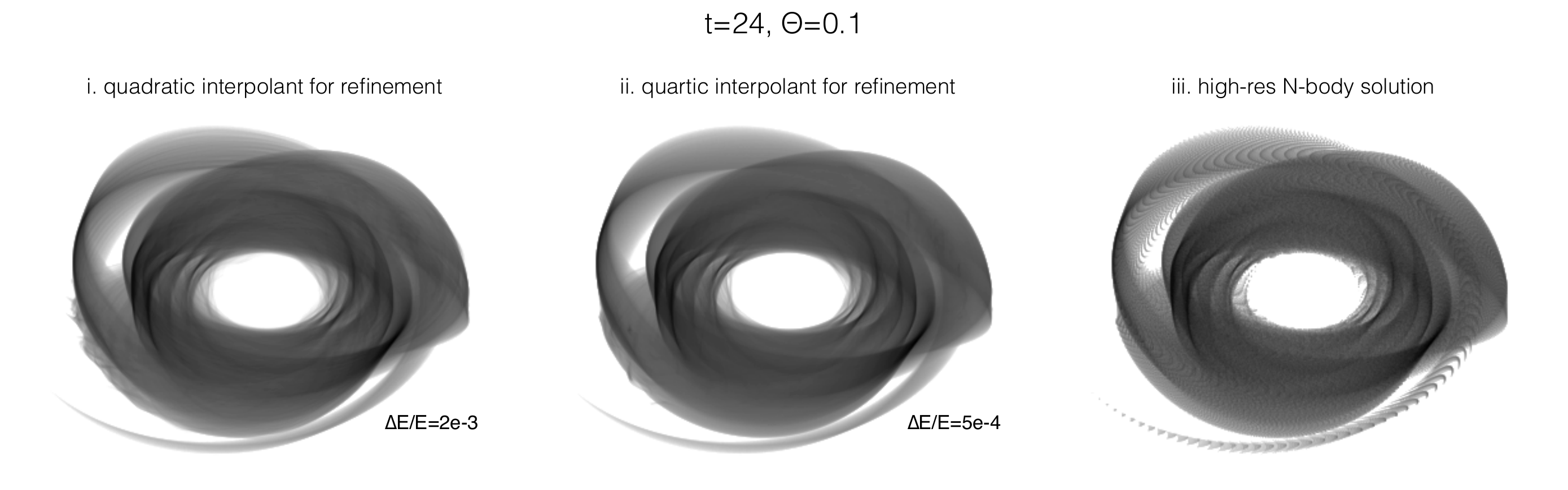}
\end{center}
\caption{
\label{fig:fixed_potential_ref_images2} The mixing cube at a much later time $t=24$ for the tri-quadratic scheme but using different orders of interpolation when inserting new vertices triggered when $\theta>0.1$. The left panel shows the final mass distribution if new vertices are inserted with the same order as the element (i.e. tri-quadratic), while the middle panel shows the result if vertices are inserted by combining 8 elements and using a tri-quartic interpolant. The right panel shows the distribution of the 2 million $N$-body particles at $t=24$. The relative error in energy conservation using the tri-quadratic interpolant is $0.2$ per cent at $t=24$, while it is $0.05$ per cent for the tri-quartic interpolant. See text for detailed discussion.
 }
\end{figure*}

\begin{figure} 
\begin{center}
\includegraphics[width=0.35\textwidth]{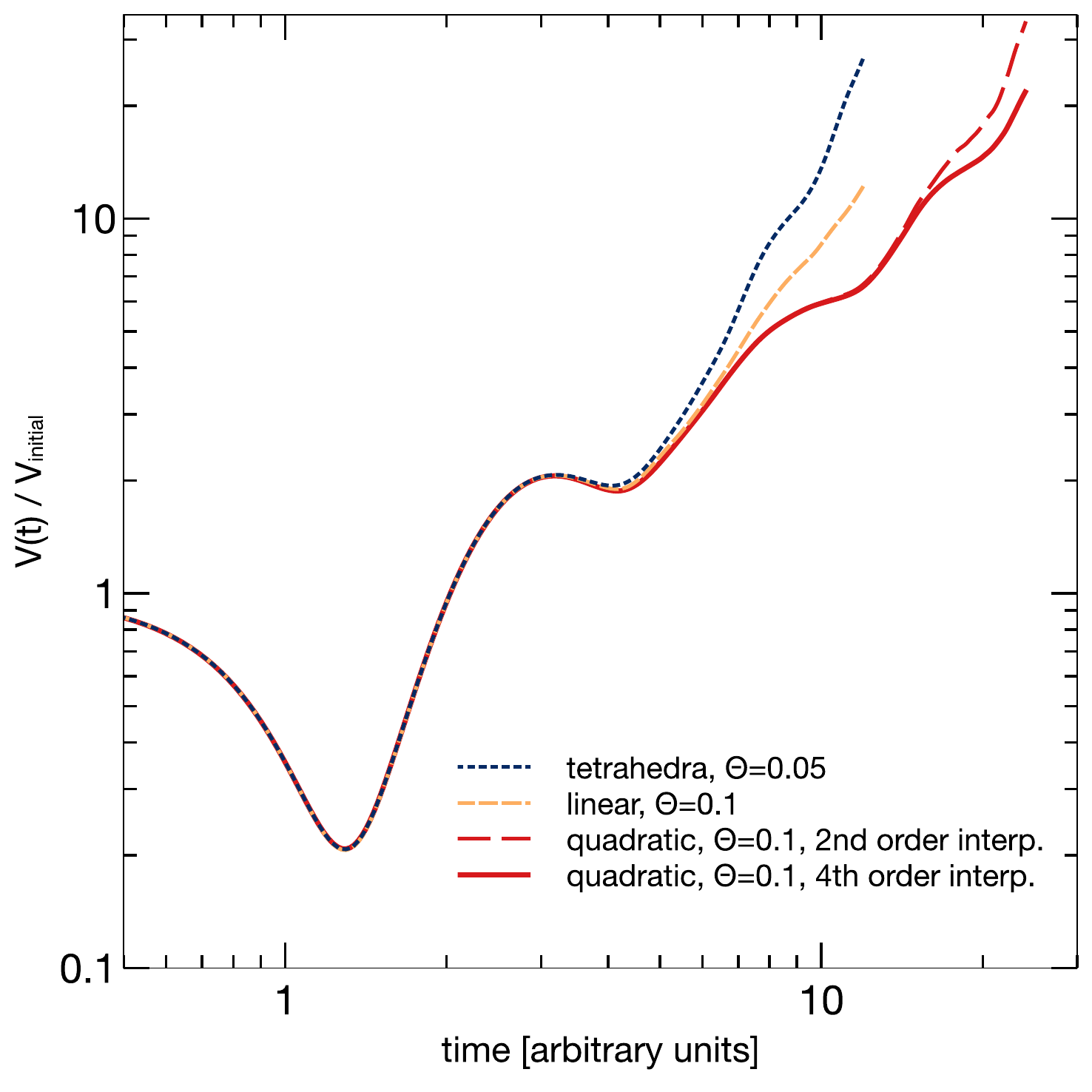}
\end{center}
\caption{
\label{fig:fixed_potential_volume} Volume over time of the mixing cube. The curves for tetrahedra (dotted blue) and tri-linear elements (yellow short-dashed) are identical to those in Figure~\ref{fig:fixed_potential_noref} but shown over a longer time interval. In addition, we use 2nd order (red long-dashed) or 4th order (solid red) tri-polynomials when inserting new Lagrangian vertices.
 }
\end{figure}

We push this observation even further and will next use a tri-quartic, i.e. 4th order interpolant over 8 tri-quadratic elements when inserting new vertices. We still represent the elements only by tri-quadratic maps. This allows to improve the quality of the solution even further. In Figure~\ref{fig:fixed_potential_ref_images2}, we show the mass distribution of the cube at a twice later time than before, i.e. at $t=24$. By this time, the inner regions have already mixed very substantially. In all cases we use a refinement threshold of $\theta=0.1$. The left panel shows tri-quadratic elements with new vertices inserted using a second order interpolant, while the middle panel uses a fourth order interpolant and the right panel shows the $N$-body reference solution. Using the fourth order interpolation improves energy conservation by a factor of 4 and once again improves the solution. Furthermore, at $t=24$, using the second order vertex insertion leads to $\sim750'000$ vertices in total, fourth order insertion requires only $\sim500'000$, which is only $1/4$ of the particles of the $N$-body reference run. The fourth order interpolation also significantly improves the evolution of the sheet volume (see Figure~\ref{fig:fixed_potential_volume}), where noise due to the second order interpolant makes the sheet grow faster than it should at very late times. 

We can conclude from the analysis in this section that adaptive refinement, especially when combined with high order schemes for mass assignment and refinement can make the Lagrangian phase space element approach energy conserving while overcoming the inherent graininess of the $N$-body method with significantly {\em fewer} active particles.


\section{Testing Self-gravitating Lagrangian Elements}
\label{sec:testing_gravity}
We have seen in the previous section that adaptive refinement can exquisitely track the evolution of the phase space sheet even in situations of strong mixing. This test was however performed in an external potential meaning that local errors in the solution did not feed back into the global solution. We will thus investigate in the subsequent sections the performance of the adaptively refined elements when self-gravity is present.

\subsection{Test 1: One-dimensional Plane Wave collapse (axis-aligned)}
\label{sec:test_1d}

\begin{figure} 
\begin{center}
\includegraphics[width=0.4\textwidth]{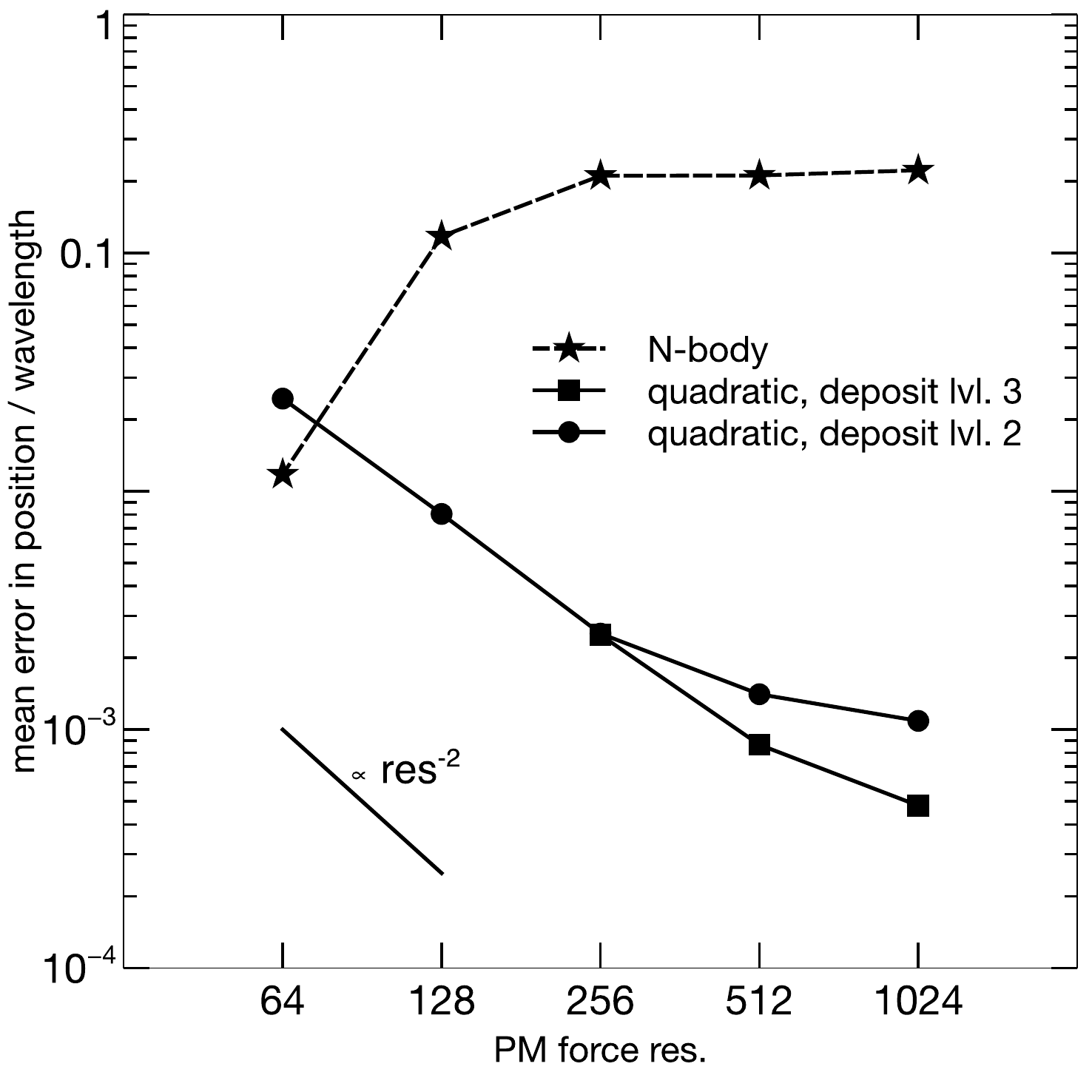}
\end{center}
\caption{\label{fig:planewave_convergence}
Error in particle displacement in the plane wave collapse problem at shell-crossing as a function of PM force resolution for $N$-body (stars) and quadratic elements. Both methods use the same number of particles and supporting points ($32^3$, which results in $16^3$ quadratic elements). The mass distribution of the quadratic elements is sampled at two different resolutions (squares and circles). See \ref{sec:test_1d} for a discussion.
}
\end{figure}

We will only briefly consider the axis-aligned one-dimensional plane wave collapse here, before we move on to computationally more challenging problems. This problem consists in solving the self-gravitating collapse, in an Einstein-de~Sitter cosmology, of a one-dimensional potential perturbation
\begin{equation}
\phi(\mathbf{q}) = A\,\cos(\mathbf{k}_p \cdot \mathbf{q}),
\end{equation}
where $\mathbf{k}_p$ is parallel to one of the cartesian axes of the computational volume, $\left\|\mathbf{k}_p\right\| = 2\pi/L$ and $A$ is chosen so that shell crossing occurs at an expansion factor of $a_{\rm cross}$. 

For the plane wave, the solution (and thus particle position, velocity and acceleration) can be calculated analytically before shell-crossing occurs \citep{Zeldovich1970} but the analytic solution, of course, relies on ``unsmoothed'' gravity. In the case that force softening goes to zero (and thus the PM resolution to infinity), the analytic solution should be approached. We thus present a plot identical to what is shown in the top panel of Figure~7 of HAK13 in Figure~\ref{fig:planewave_convergence}: the position error of flow tracers at the time of shell-crossing as a function of the force resolution while keeping the number of flow tracers constant. The solution should converge to a sharper and sharper caustic with increasing force resolution. As demonstrated by HAK13, $N$-body does not converge at fixed particle number. Here, we show the results for standard $N$-body and for tri-quadratic phase space elements {\em without adaptive refinement}. Unlike HAK13, where no true convergence at fixed tracer particle number was seen, we see that the recursive mass deposit improves the situation substantially, leading to {\em close to second order convergence  to the analytical solution with a constant number of tracer particles!} For a given deposit recursion level, the error w.r.t. the analytical solution always asymptotes to a fixed value. To get true second order convergence, either exact deposits have to be employed (which have been implemented e.g. for tetrahedra by \citealt{Powell:2014}), or an optimally chosen deposit level for a given force resolution has to be employed.


\subsection{Test 2: One-dimensional Plane Wave collapse (oblique)}
\label{sec:oblique_wave}
\begin{figure} 
\begin{center}
\includegraphics[width=0.45\textwidth]{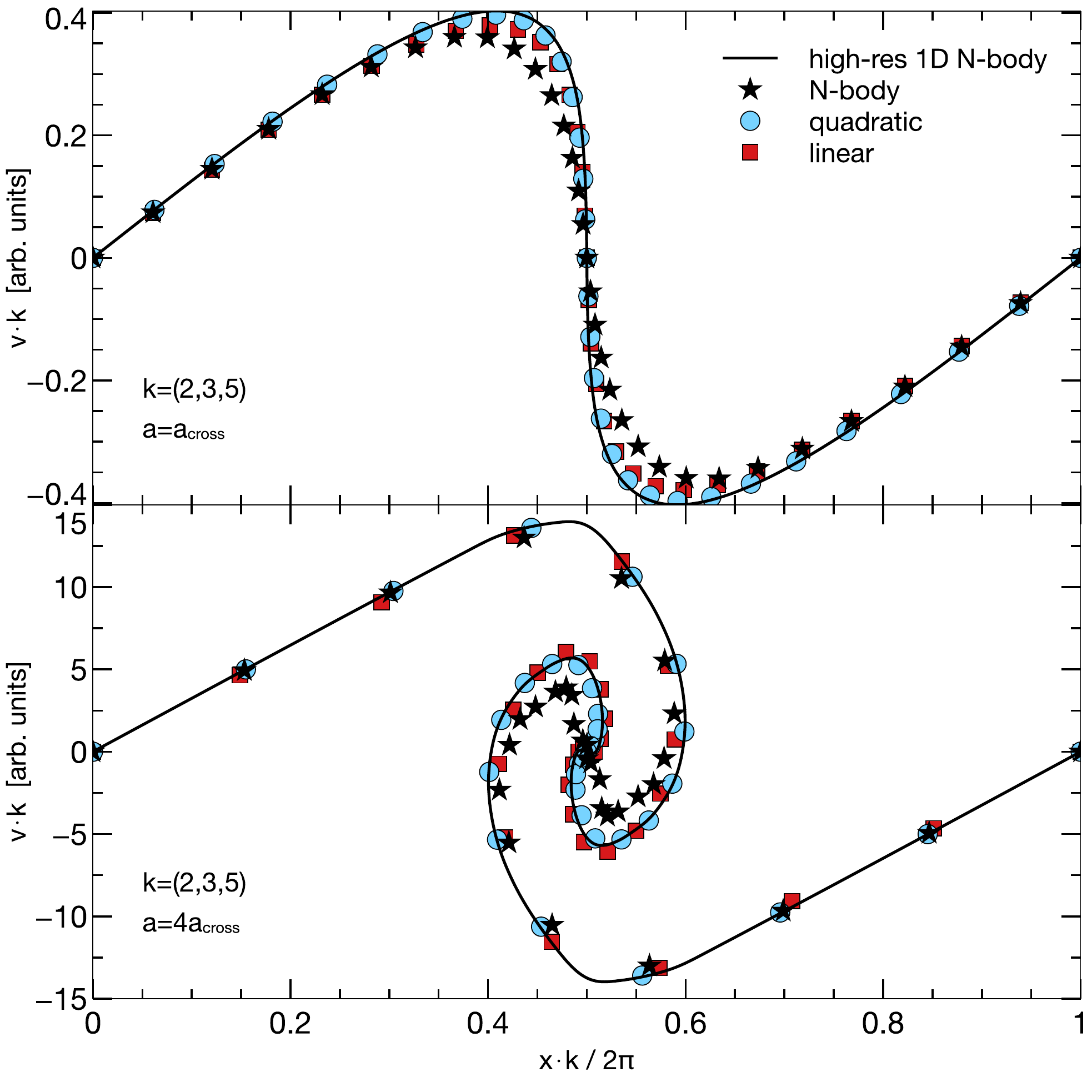}
\end{center}
\caption{
\label{fig:oblique_wave} Collapse of a plane wave not aligned with the initial particle lattice. The phase space projected along the wave vector is shown at shell-crossing for standard $N$-body (black stars), the tri-linear scheme (red squares) and the tri-quadratic scheme (blue circles). The black line indicates a very high-resolution 1D solution at identical force-resolution. In the top panel, the wave is shown at shell-crossing ($a=a_c$), in the bottom panel at $a=4a_c$. Note that points have been re-mapped to the unit interval of the wave mode so that points that appear as neighbours in phase-space are not neighbours but spread throughout the computational box.
 }
\end{figure}

We next consider a single plane wave that is not aligned with the Cartesian axes, but has a wave vector  $\mathbf{k}_p=(2,3,5)k_f$, where $k_f=2\pi/L$ is the fundamental mode of the simulation box. This is a very hard problem for particle methods since the one-dimensional wave motion has to be preserved in three dimension \citep[cf.][]{Melott1997}. The particular choice of $\mathbf{k}_p$ leads to the particles or flow tracers being folded onto themselves at rationals of the box-length. In Figure~\ref{fig:oblique_wave}, we show the phase space along the wave vector for the new phase space element method (using tri-linear and tri-quadratic elements {\em without adaptive refinement}) as well as the standard N-body method. We employed a force resolution of $256^3$ and $32^3$ tracer/N-body particles. The mass distribution of the phase space elements was sampled at two deposit levels. We compare the results to a high-resolution one-dimensional solution (using $16,000$ particles) at the same force resolution at two times corresponding to $a=a_{\rm cross}$ (top panel) as well as $a=4a_{\rm cross}$ (bottom panel). 

\begin{figure*} 
\begin{center}
\includegraphics[width=\textwidth]{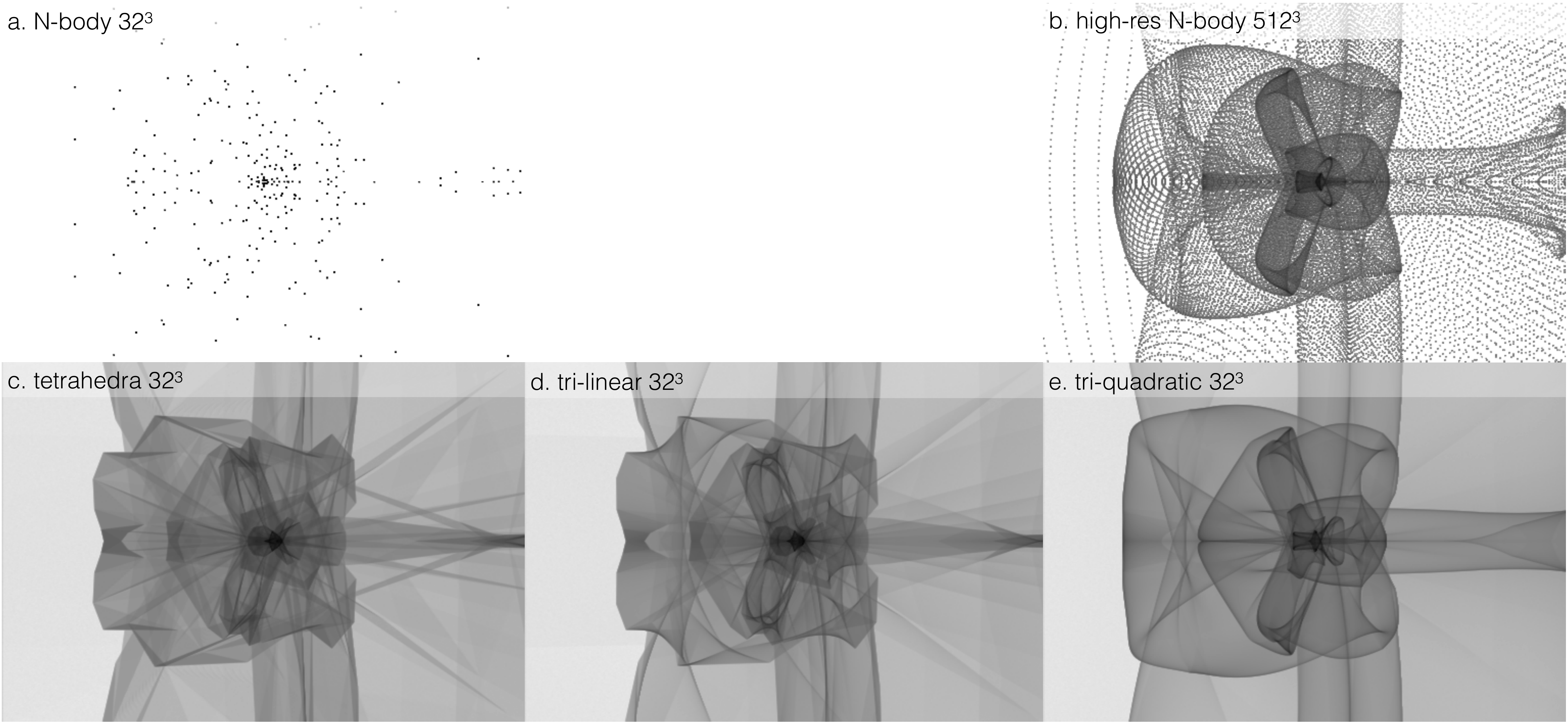}
\end{center}
\caption{
\label{fig:ripple_wave} Impact of the order of the Lagrangian elements on the solution {\em without refinement}: the ``ripple wave'' test problem using $32^3$ flow tracers and PM force resolution of $256^3$ cells for standard $N-body$ using $32^3$ particles (panel a), using $8\times32^3$ tetrahedral elements (panel c), $32^3$ tri-linear elements (panel d) and $16^3$ tri-quadratic elements (panel e). These solutions are compared to a high-resolution $N$-body run employing $512^3$ particles (top right, panel b) at the same force resolution. The degrees of freedom of the system in panels a,c,d and e are thus the same, while panel b has $16^3$ more. A rendering of panel a, using tri-quadratic elements a posteriori is shown in Fig.~\ref{fig:ripple_wave_triquad}.
 }
\end{figure*}

The results are consistent with what HAK13 found (the results in their Figure~11 are however shown at somewhat later times) in that the $N$-body solution differs widely from the true one-dimensional solution. Here we see that this discrepancy already exists at shell-crossing, but becomes most severe at later times, where substantial velocities outside of the plane of symmetry are found to break momentum conservation inside of the symmetry plane. Quite differently for the tri-linear and tri-quadratic phase space elements. Using the same number of $32^3$ tracer particles as the $N$-body solution, they have substantially smaller errors at shell crossing and trace the correct solution still very well at much later times when $a=4a_{\rm cross}$, in fact, the tri-quadratic solution is right on top of the high-resolution solution. This is even more remarkably since the way we plot the solution makes it appear that the resolution is higher than it actually is: points that appear as neighbours in phase-space in this plot are not actually neighbours in Lagrangian space. They are points at the respective phase of the wave but at rather different locations on the Lagrangian manifold and therefore spread out throughout the entire simulation box.

\subsection{Test 3: two-dimensional perturbed plane wave collapse}
\label{sec:test_2d}
\begin{figure}
\begin{center}
\includegraphics[width=0.8\columnwidth]{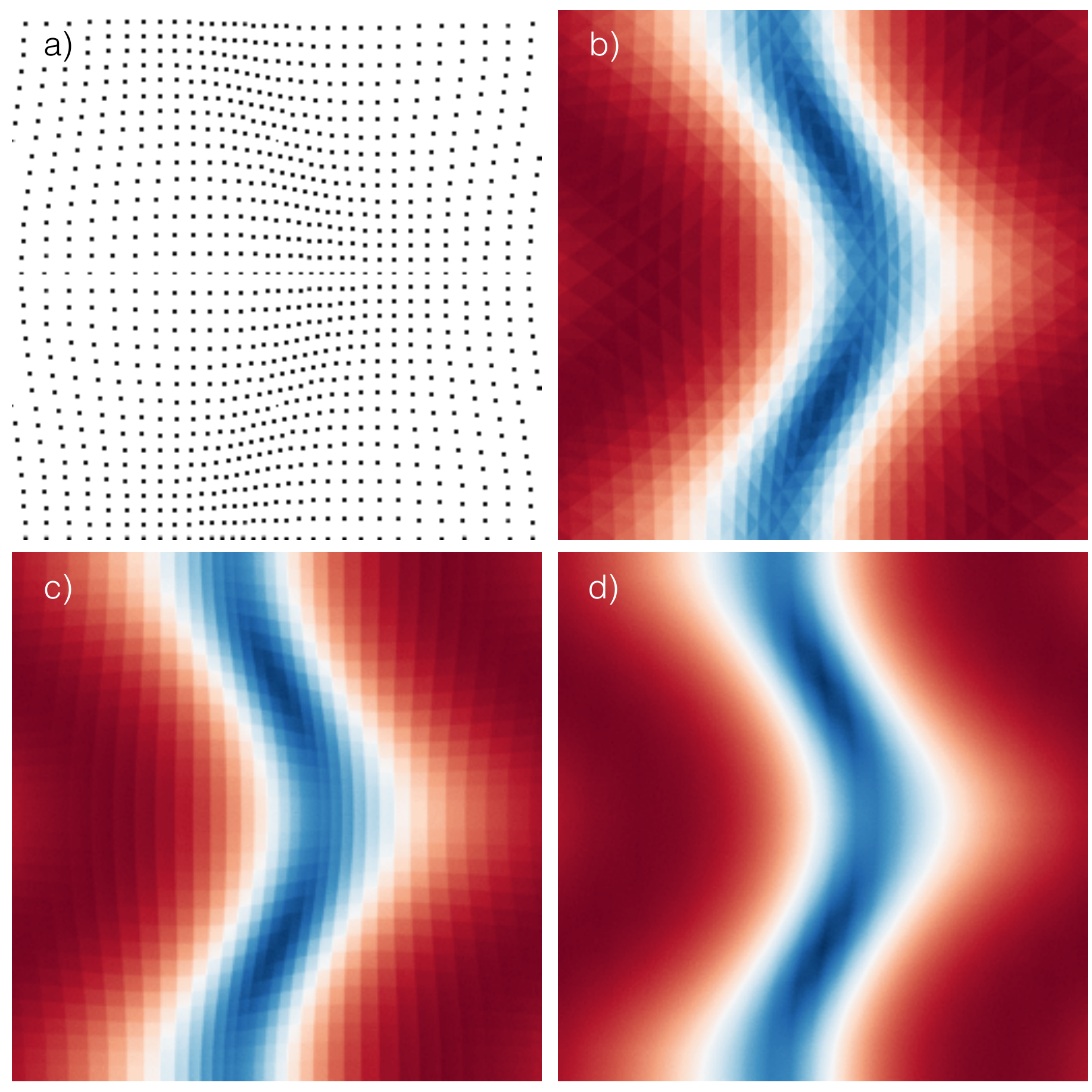}
\end{center}
\caption{
\label{fig:ripple_ics}The initial conditions for the ``ripple-wave'' test problem (cf. Sec.~\ref{sec:oblique_wave}). Shown are the particle locations (panel a), the density field using the tetrahedral phase space elements (panel b), using tri-linear elements (panel c) and using tri-quadratic elements (panel d). The linear elements are discontinuous at element boundaries, while the quadratic is continuous.
}

\end{figure}

In a final test problem, we employ the same two-dimensional test as in \cite{Hahn2013}:
the anti-symmetrically perturbed wave described  by \cite{Valinia1997}. Compared to the
plane wave collapse above, this problem introduces anisotropic collapse as
well as mixing and is thus a significantly harder problem. In this test problem,
the Lagrangian gravitational potential is given by that of a plane wave in x-dimension 
with a sinusoidal phase perturbation in the y-dimension
\begin{equation}
\phi(\mathbf{x}) = \bar{\phi} \cos\left(k_p \left[ x + \epsilon_a \frac{k_p}{k_a^2} \cos k_a y \right]\right).
\end{equation}
The initial particle positions and velocities have been obtained using the Zel'dovich approximation at $z=19$. We adopt 
$k_p = 2\pi / L$, $k_a = 2\pi/L$ and $\epsilon_a=0.5$, where $L=10\,h^{-1}{\rm Mpc}$ is the size of the simulation box (note that this is different from the parameters chosen in HAK13), and $\bar{\phi}$ is  chosen so that first shell crossing occurs at an expansion factor of $a_c=1/7.7\simeq0.13$. We run this problem in an Einstein-de~Sitter cosmology and discuss the results at $a=0.5$.  In all cases, unless otherwise specified, we use initial conditions for $32^3$ particles and employ a force resolution of $256^3$ PM cells. All solutions are run in three dimensions, effectively replicating the problem along the $z$-axis. The initial density fields for this test problem using the particle representation as well as the elements of various orders can be seen in Figure~\ref{fig:ripple_ics}. As already discussed in Section~\ref{sec:densities}, the tetrahedral elements provide a piecewise constant, while the tri-quadratic provide a smoother continuous density field.

In Figure~\ref{fig:ripple_wave}, we present the solution for this test problem using a constant number of elements of increasing order, from top to bottom: $N$-body, then tetrahedra, trilinear elements, and tri-quadratic elements, and a reference high-resolution $N$-body run using $512^3$ particles at the same PM resolution. We see that all Lagrangian element methods recover a roughly identical structure that is only barely inferred visually from the low-resolution $N$-body solution. Furthermore, we clearly see how the increasing order of the elements improves the smoothness of the resulting density field (as shown in Section~\ref{sec:densities}). The most important results to be taken from this figure are the central density of the clump and the position of the caustics. HAK13 have found that the self-gravitating tetrahedra can lead to a biased central density and also bias in the positions of caustics. We see these shortcomings reproduced in this figure for both tetrahedra and tri-linear elements. The only advantage of the tri-linear elements is that the projected density is somewhat smoother. As expected, both the central density and the positions of caustics (compared to the high-resolution $N$-body solution) are, at least visually, significantly improved when the tri-quadratic elements are used.

\begin{figure} 
\begin{center}
\includegraphics[width=0.75\columnwidth]{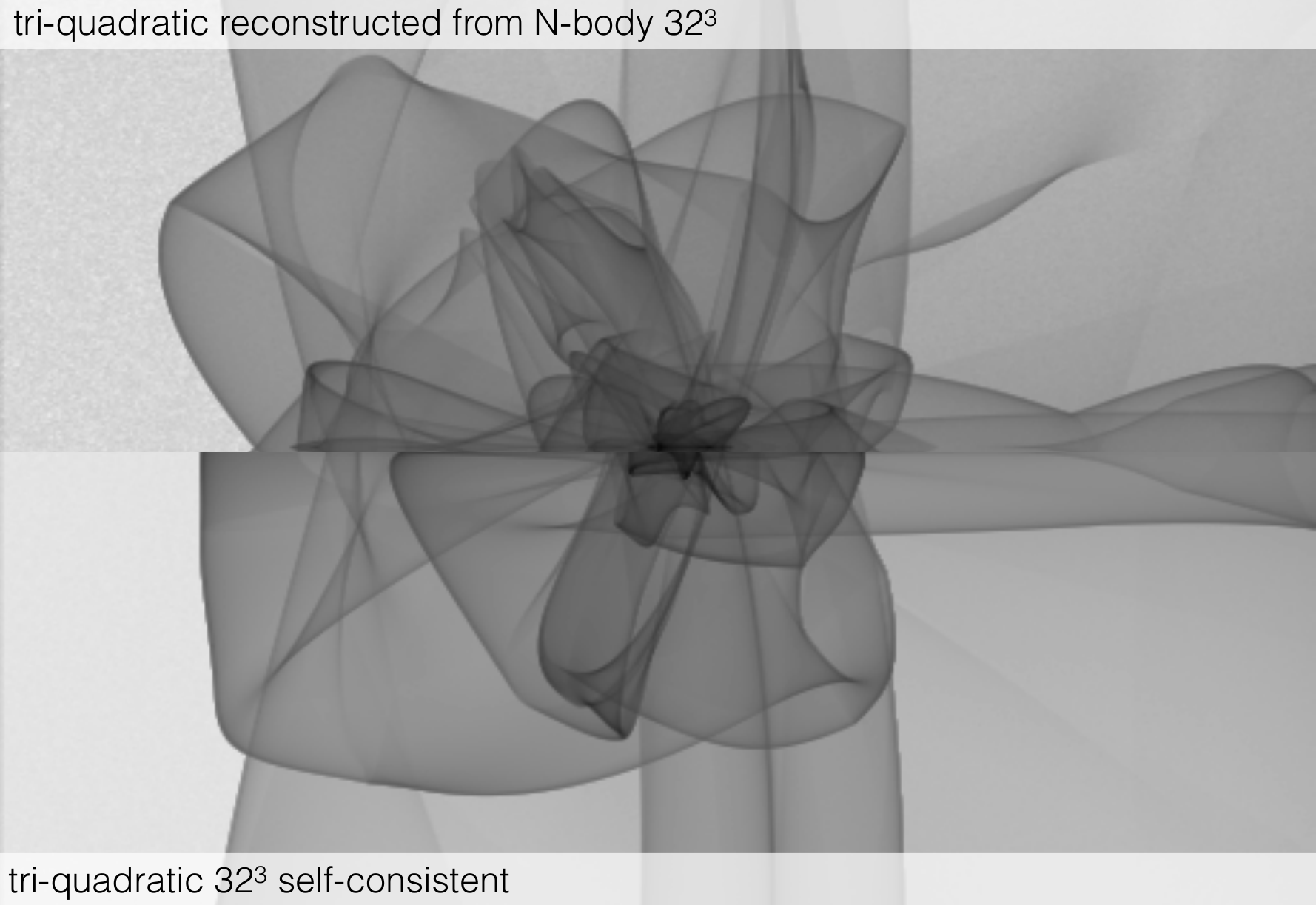}
\end{center}
\caption{Comparison between a reconstruction of the tri-quadratic density field from the $32^2$ standard $N$-body run (top half-panel) and the self-consistent evolution of the tri-quadratic elements (bottom half-panel). One clearly sees that $N$-body particle noise significantly perturbs the solution, in particular, caustics are not persistent.
\label{fig:ripple_wave_triquad} 
}
\end{figure}

Clearly, the density field inferred from the tri-quadratic elements is significantly smoother than that from the linear element types. One might thus wonder whether $N$-body actually does well but looks bad in comparison. To illustrate that this is clearly not the case, in Fig.~\ref{fig:ripple_wave_triquad}, we show the density field using tri-quadratic elements, determined a posteriori from the evolved low-resolution $N$-body solution (top half-panel), in comparison to the self-consistent tri-quadratic element solution (bottom half-panel). Although the quadratic elements provide a smooth field, the $N$-body estimate show a very noisy solution, where ``noisy'' now means that caustics are significantly perturbed compared to the much smoother self-consistent solution. The smooth field is thus clearly not an artefact of the rendering, but demonstrates convincingly the superiority of the Lagrangian elements over the $N$-body solution!



\begin{figure*} 
\begin{center}
\includegraphics[width=\textwidth]{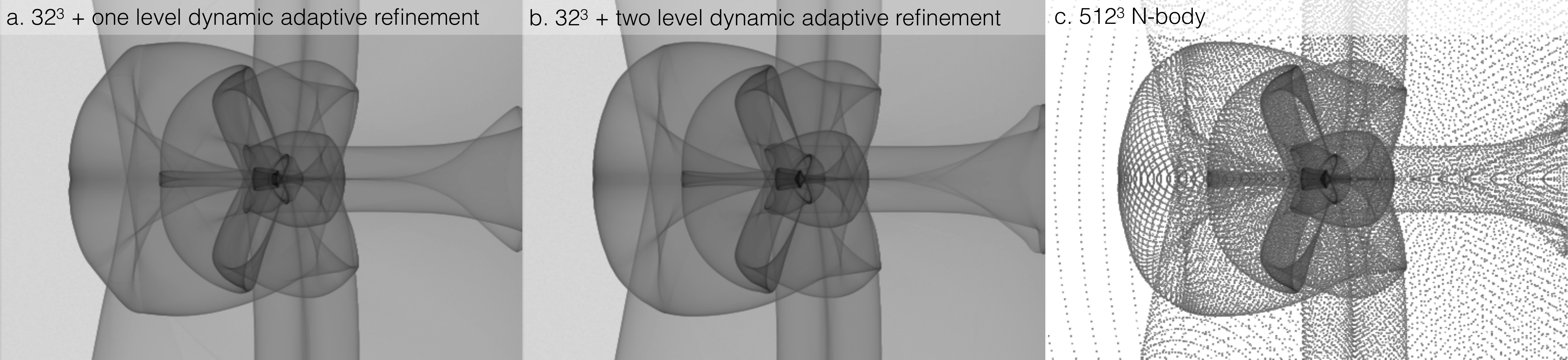}
\end{center}
\caption{The ripple wave collapse test with dynamic adaptive refinement. The $32^3$ runs use the same initial conditions as in Fig.~\ref{fig:ripple_wave}, tri-quadratic elements and one (left, panel a), and two (middle, panel b) of dynamic adaptive refinement. The right panel shows again the solution of a high-resolution $N$-body run using $512^3$ particles at the same $256^3$ PM force resolution. One clearly sees how adding more supporting points approaches the high-resolution $N$-body solution. Still, the left two panels have significantly fewer degrees of freedom than the $N$-body run.
\label{fig:ripple_wave_refine} 
}
\end{figure*}

Naturally, the big remaining question is whether dynamical adaptive refinement in such a self-gravitating system with mixing converges to the right solution. We show the solution using refinement in Figure~\ref{fig:ripple_wave_refine}, comparing once more against the $512^3$ particle high-res $N$-body solution at the same force resolution. We only consider the tri-quadratic elements in this case, although the linear elements also perform reasonably well. We started with the same $32^3$ initial conditions as in the fixed resolution test shown in Figure~\ref{fig:ripple_wave}, but now employed the force refinement criterion with a threshold of 0.1 to dynamically split elements if required (the results using velocity refinement are however not significantly different). The solution allowing for one additional level of refinement is shown in the top panel, the one for two levels in the middle panel, and the reference $N$-body solution at the bottom. Rather strikingly, the solutions quickly converge to the reference solution in the exact shape and position of caustics. Already with one additional level, the central density of the clump is comparable to the reference solution. We do not perform a more quantitative solution of these toy problems but let the images speak for themselves and perform a quantitative convergence study of refinement in the next section, where we apply the Lagrangian  element method to cosmological structure formation.

\section{A First Application: Cosmological Simulation of a Warm DM Universe}
\label{sec:cosmo}

We now apply our Lagrangian phase space element method to a cosmological problem. We simulate the
gravitational evolution of a L=$20$ Mpc/h cube in a universe where dark matter
is made of warm particles of mass $m_{\rm dm} = 250\,{\rm eV}$, leading to a small-scale cut-off
in the density perturbation spectrum.

\begin{figure*}
\includegraphics[width=\columnwidth]{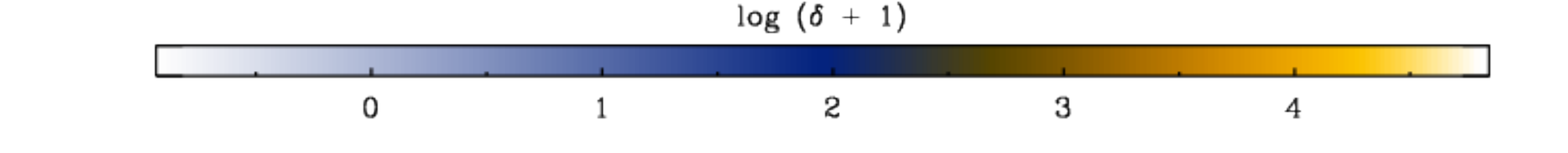}
\includegraphics[width=\columnwidth]{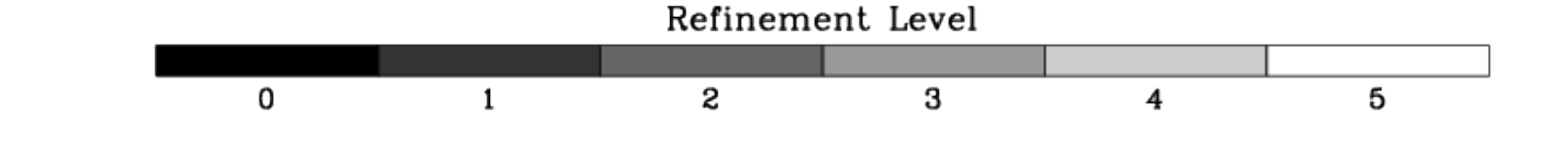}
\includegraphics[width=\columnwidth]{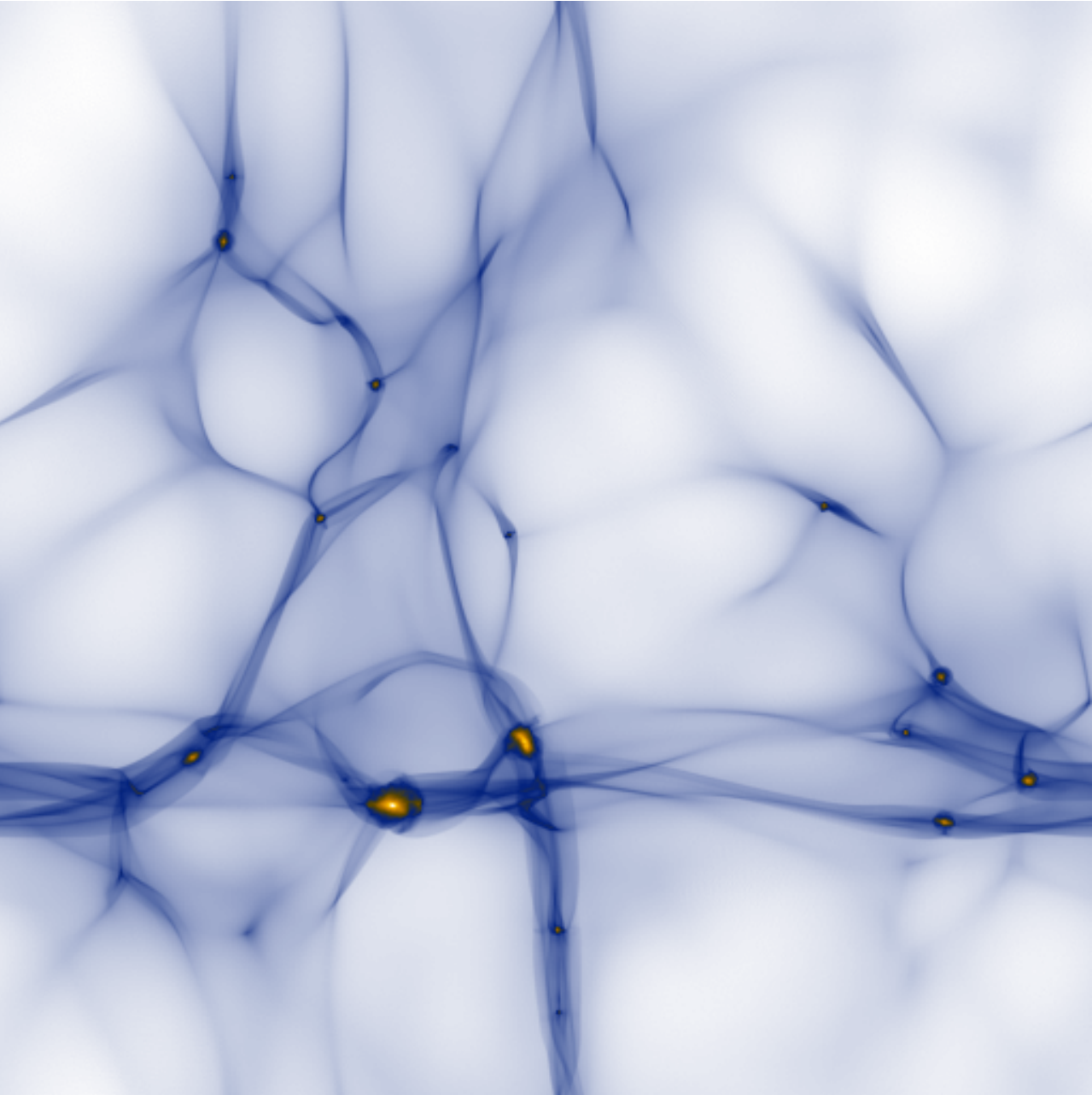}
\includegraphics[width=\columnwidth]{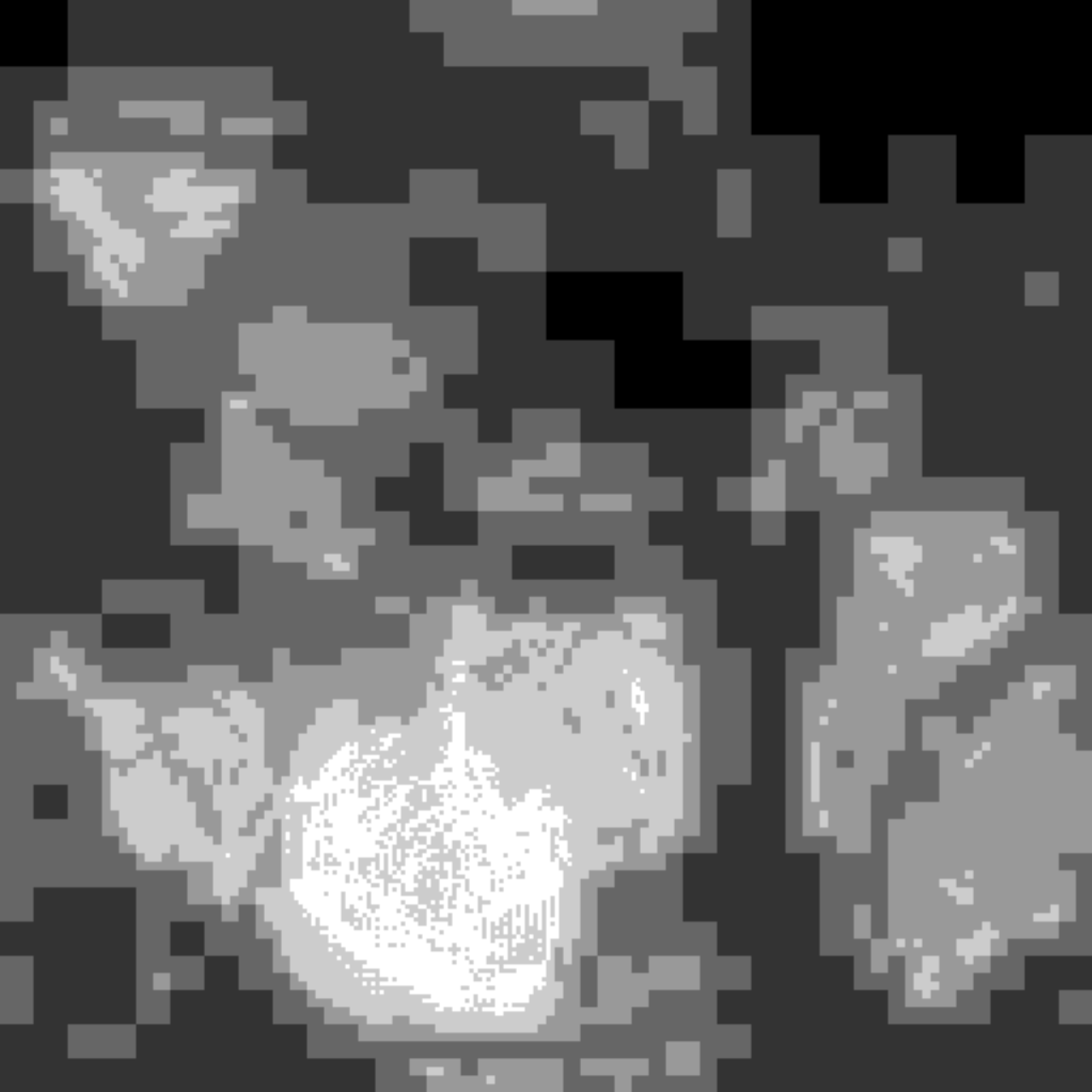}
\caption{A Warm DM universe at $z=1$ simulated employing the adaptively refined
    phase-space element method presented in this paper. We use here the tri-quadratic
    elements. The left panel
    shows the logarithm of the projected mass density field. The right panel shows
    a map of the final refinement level reached by Lagrangian patches; in black
    regions that did not trigger our refinement threshold, and in white those that
    reached the maximum refinement level allowed in this simulation.
    \label{fig:full}
}
\end{figure*}

The cosmological parameters we employ correspond to those favoured by the WMAP7
data release measurements \citep{Komatsu2011}. Explicitly: $\Omega_m = 0.276$,
$\Omega_{\Lambda} = 0.724$, $\Omega_{b} = 0.045$, $h = 0.703$, $\sigma_8 =
0.811$ and spectral index $n_s = 0.96$. We generate our initial conditions
using the MUSIC code \citep{HahnAbel2011}, with an input linear density
power spectrum given by the fitting formulae of \cite{EisensteinHu1999} and a
small-scale cutoff that mimicks the effects of a warm dark matter particle
\citep[see][for more details]{Bode2001,Angulo2013}.

We follow the evolution of $16^3$ Lagrangian patches, each one of mass
$1.49\times10^{11}\,\Mass$, which are created from a set of $64^3$ particles.
Gravitational forces are computed using a PM method with a grid of $512^3$
cells, which implies that forces are effectively softened below $80$ kpc/h. We
note that the dark matter particle mass, cosmological parameters, simulation
particle mass and force resolution match those of \cite{Angulo2013}.

In the following, we compare the results of simulating the above configuration
with $5$ different methods; i) a standard $N$-body method, with $64^3$ particles, ii) the tetrahedral
method of \cite{Hahn2013} and \cite{Angulo2013}, with a total of $8\times32^3$ tetrahedra,
iii) $32^3$ tri-linear elements, iv) $16^3$ tri-quadratic elements, and v) $16^3$ tri-quadratic elements with
1 or 3 levels of dynamic adaptive refinement allowed. They {\em all} employ numerically identical initial conditions
and simply group particles together as vertices to define the elements.

The simulations that feature our Lagrangian element method were run with $2$ levels of
recursion in their mass deposit. This means that the density field associated
to a given Lagrangian patch was represented with $42496$ bodies, summing up to
$174063616$ in the entire simulation volume. The simulation that was allowed to
refine up to 3 levels created $302737$ new resolution elements by $z=1$. At
this redshift, we find only $18$ patches that have not triggered refinement,
$19794$, $76885$, and $206040$ at refinement levels 1,2, and 3, respectively.
We recall that at the highest refinement levels, particle have a mass $8^3$
times smaller, i.e $m_p = 2.8\times10^{8}\,\Mass$.

The evolution was simulated from $z=63$ to $z=1$ using $192$ global time-steps.
As could have been expected, the computational costs of the different runs
differ largely. The standard $N$-body run took $81$ CPU mins. The tetrahedra,
linear and quadratic runs took $144$, $305$, $302$ CPU mins, respectively. The
run with refinement took$ 257$ CPU hours. Out of this time, $70\%$ was spent in
the mapping of mass carriers onto the PM grid, and $20\%$ in the on-the-fly
generation of mass carriers.

In Fig.~\ref{fig:full}, we illustrate the performance of our adaptively-refined
Lagrangian method. In the left panel, we show the final density field in
Eulerian coordinates at $z=1$. The high level of detail that our
method provides about the topology of the large scale structure in WDM is clearly obvious. One should keep in mind that this is
while using the equivalent of evolving only $64^3$ points. We highlight that the image displayed is
not a smoothed rendering of our patches, but it is the actual density field
used inside our simulation code to gravitationally evolve the dark matter
fluid. On the right hand side image, we show a map of the highest refinement
level reached by different regions in Lagrangian coordinates (shown as the maximum along the line-of-sight).  It is clear that
the refinement level correlates with the Lagrangian counterpart of dense
filaments and haloes, and that the material infalling into the largest haloes
will lead to the largest amounts of refinement.

\begin{figure}
\includegraphics[width=\columnwidth]{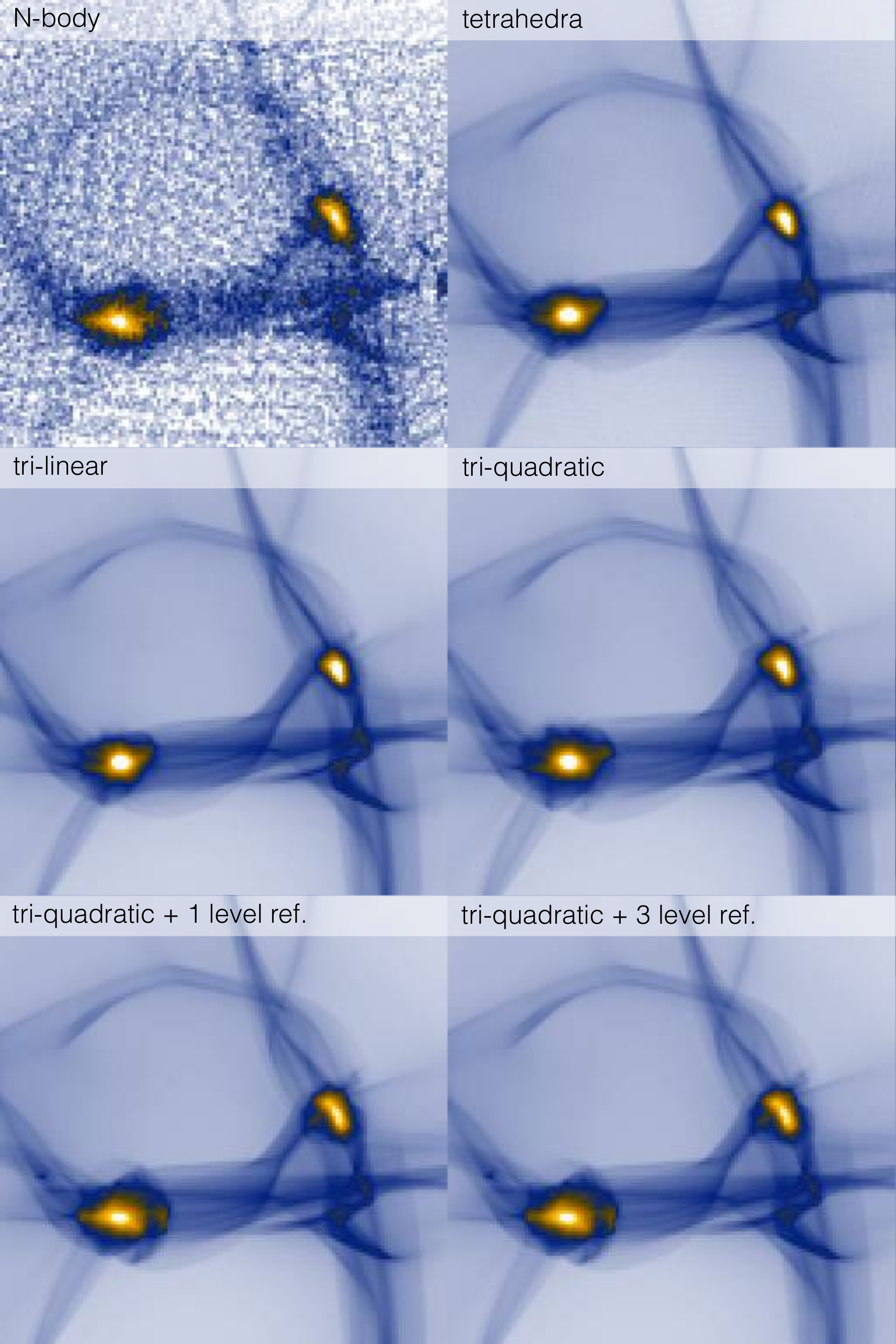}
\caption{The projected mass field in a region of dimensions $20\times5$ Mpc/h
    of our WDM simulations at $z=1$. Each panel displays the field predicted by
    different methods, as indicated in the top left corner. The image
    corresponds to the logarithm of the actual density field used to computed
    gravitational forces trough the Poisson equation. Note that the colour
    scale is identical in all panels. The top four panels have the same number of
    degrees of freedom (particles), while in the bottom two panels, the degrees
    of freedom increase over time to capture mixing in the haloes.
    \label{fig:wdm}
}
\end{figure}

After this first impression of the full performance of our approach, we now qualitatively compare our different runs. In Fig.\ref{fig:wdm} we show
the density field at $z=1$ for a $5\times5$ Mpc/h region centred on the most
massive haloes present in the box.  We present the same image for all the cases mentioned above, as indicated by
the legend. As in the previous image, this is a not a ``visualization-purpose'' rendering of the
particle distribution, but the actual density field that sources the
gravitational potential field trough the Poisson equation in our gravitational evolution.

It is readily visible that all methods coincide in their predicted large scale
structure: filaments, halos and voids coincide in their location and extent.
However, there are also clear differences among the methods. Perhaps the most
striking difference is the evident granularity of the standard $N$-body method
and how filaments are broken into clumps in this case. In contrast to this, all of our
Lagrangian methods provide an extremely smooth field where filaments are {\em continuous}
1D structures. Furthermore, some of these filaments are embedded inside larger
filaments which shows very clear caustics, none of these features can be even
distinguished in the standard $N$-body case. Again, we want to emphasise that the top four
panels in this figure employ the exact same number of ``particles''.

When comparing the runs using the new method, we can see that they most strongly differ in the internal
structure of the massive haloes present in the image. The cases employing
tetrahedra, tri-linear and even tri-quadratic elements all display very concentrated
and round halos (with the tri-quadratic elements improving slightly). This is a consequence of the biases extensively discussed
throughout this paper. The situation is clearly different when adaptive refinement
is enabled and the halos appear less dense and more triaxial, very much in agreement
with the standard $N$-body case.

\begin{figure}
\includegraphics[width=\columnwidth]{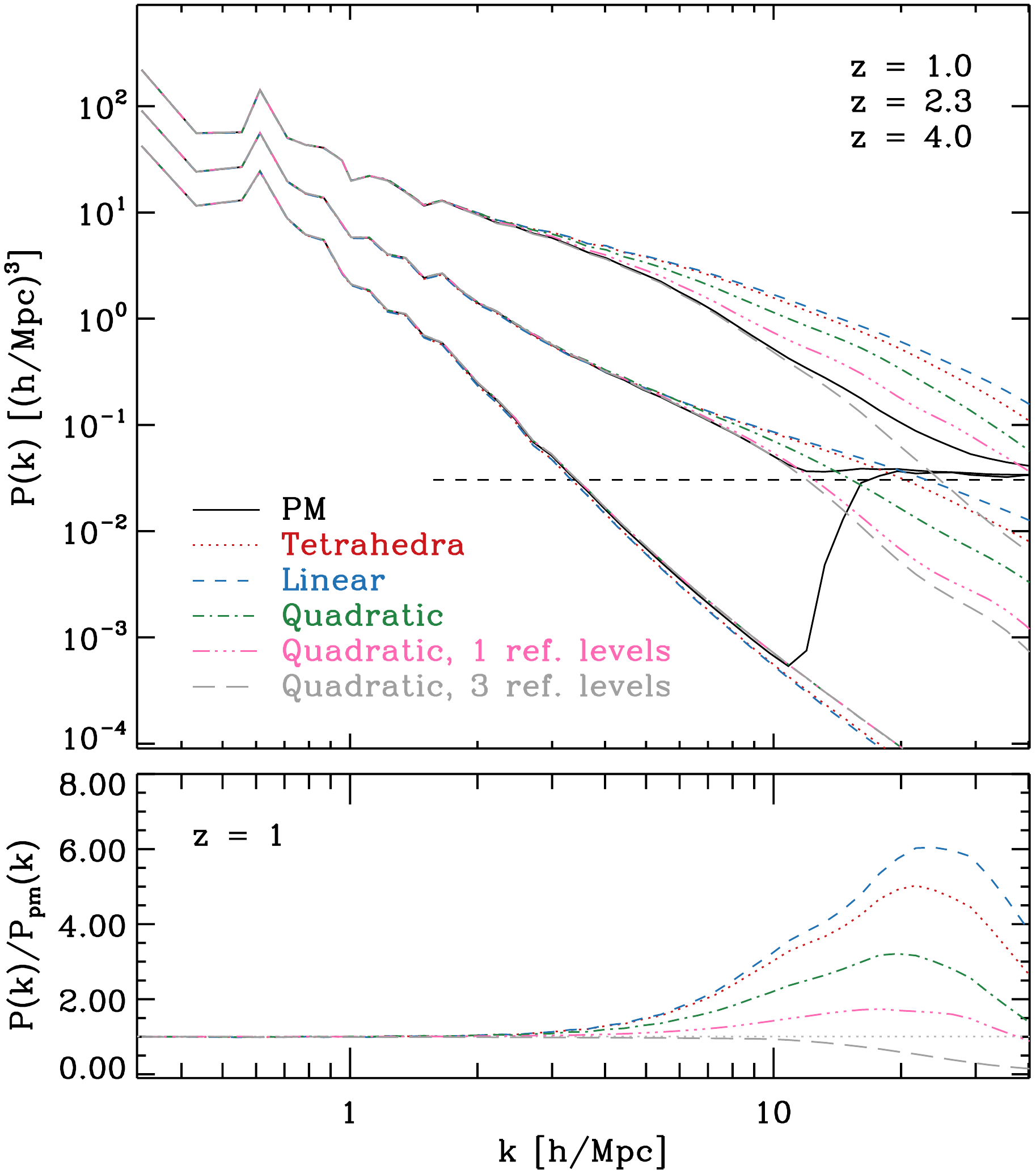}
\caption{Matter power spectra measured in our WDM simulation at three different
    redshifts, z = $4$, $2.3$, and $1$ from bottom to top. At each redshift,
    the six lines display the results using different methods to simulate the
    evolution of the dark matter density field: standard PM $N$-body (solid black), tetrahedral
    (dotted red), tri-linear (dashed blue), and tri-quadratic elements (dash-dotted green), as well as
    tri-quadratic with one (dash-dot-dot, pink) and three dynamic levels of refinement allowed (long dashed grey), as indicated by the legend.
    The Poisson shot noise at $z=1$ is indicated by the
    horizontal dashed line. In the bottom panel we display the ratio of the
    measured spectra to the spectra computed by the standard $N$-body method at $z=1$.
    For the four non-refined methods, the degrees of freedom (particles) are identical and
    and constant and they only differ in the order of the mass projection.
    \label{fig:wdm_pk}
}
\end{figure}

In Fig. \ref{fig:wdm_pk} we quantify these differences by comparing the density
power spectrum among our runs. This figure shows three groups of curves, from
bottom to top: $z=4$, $2.3$ and $1$, respectively. Within each set, coloured
lines show the results of the 6 cases discussed above and indicated by the
figure legend. At high redshift, we see that all methods are almost
indistinguishable on large scales. On small scales, the standard $N$-body case
strongly deviates from the rest due to the dominating effect of the primordial
lattice, which drives the measured power spectrum to the Poisson shot noise
level on scales smaller than the mean inter-particle separation. At $z=2.3$ the
differences between methods start to appear at $k > 5$ h/Mpc. However, the
contribution of shot-noise in the $N$-body case is not negligible on such
scales. One can however already see that the power spectrum of the Lagrangian
element methods without refinement is biased high, while those with refinement
are consistent with the $N$-body solution.

At $z=1$ we can then clearly compare the performance of the different methods. In agreement with the
qualitative picture provided by Fig. \ref{fig:wdm}, tetrahedral and tri-linear
elements lead to a factor of $\sim4$ more power than the $N$-body case
at $k = 10\,h^{-1}{\rm Mpc}$. Tri-quadratic elements and one level of adaptive refinement
significantly reduce the excess power, are, however, still not sufficient to resolve the phase space sheet
evolution well enough. Clearly, allowing for
more levels of dynamic refinement solves this problem. When three levels are allowed, then for $k < 10$ Mpc/h the
measured spectrum perfectly agrees with standard $N$-body case. An accurate comparison is
not possible on smaller scales since the N-body case becomes dominated again by
the Poisson noise intrinsic to its discrete nature. The adaptively refined tri-quadratic elements however probe
the density field to much smaller scales with no such noise term, as we had seen in Figure~\ref{fig:wdm} qualitatively.


\section{Discussion and Conclusions}
\label{sec:conclusions}

Numerical simulations of the evolution of the dark matter fluid have been and
will be the backbone of physical cosmology in the late Universe. They are
essential to our understanding of the formation of structures in the Universe --
voids, filaments and clusters -- as well as of the evolution and formation of
galaxies and the distribution of dark matter around our Milky Way. 
In addition, they are indispensable in the correct and accurate
interpretation of virtually all cosmological observables, and thus, they are
fundamental to almost every aspect of dark energy probes and dark matter
experiments.

Methodologically, however, all of these predictions rely on only few techniques in the mildly non-linear regime: namely perturbative approaches and effective field theories; and only a single technique in the highly-nonlinear regime in three dimensions: namely the $N$-body method. While it may appear to the outsider that there is a wide range of numerical tools available, PM, TreePM, fast-multipole $N$-body, direct N-body, AMR, these approaches only differ in how they calculate the forces between the particles, but not how they discretise the dark matter fluid \citep[see e.g.][for a recent review and discussion of the $N$-body method for both collisional and collisionless
dynamics]{Dehnen2011}. In all these methods the collisionless Vlasov-Poisson system with cold initial conditions is discretised in terms of the characteristics of $N$ massive particles. It can be shown that as $N\to\infty$, the $N$-body system approaches the Vlasov-Poisson system, but naturally these simulations are far from that limit. As we have argued in length in the Introduction, there are several known short-comings of the $N$-body approach, most notably spurious fragmentation in simulations with a cut-off scale in the initial perturbation spectrum, as well as two body scattering. Recently, \cite{Hahn2013} have presented a promising alternative to $N$-body that was shown to avoid spurious fragmentation by performing a tessellation of the $N$-body particles in Lagrangian space and then assigning the mass to the tetrahedral elements (whose connectivity is preserved throughout the simulation) instead of the vertices as in the $N$-body approach. At the same time, the method of \cite{Hahn2013} is limited by the strong constraint that the tetrahedra should describe the mass enclosed between the particles accurately. This however is a constraint that is known to be violated in regions of strong mixing, manifesting itself as a density bias in the centres of halos.

In this paper we present a more general discretisation of the phase space sheet in terms of ``Lagrangian phase space elements'' which are a generalisation of both $N$-body and the tetrahedral tessellation approach. We can summarise our results as follows:

\begin{enumerate}[(1)]

\item We propose a new ``Lagrangian phase space element'' method in Section~\ref{sec:method} that describes a piecewise mapping between Lagrangian space and Eulerian phase space. We specifically discuss tri-polynomial maps in this paper, but other functional forms are possible. The Jacobian of the map to configuration space defines a unique density for every point in the element, and the projection of all Jacobians provides a density field in Eulerian space of arbitrary regularity, determined only by the order of the chosen polynomials. Specifically we show that the tetrahedral approximation of \cite{Hahn2013} is a low order version implementing only a piecewise linear transformation between the two spaces. We consider explicitly tri-linear and tri-quadratic elements here, the latter giving a continuous density field and the latter two being able to represent caustics on a single-element level.

\item We then discuss a simple pseudo-particle discretisation of the elements. This simplifies the problem of projecting high-order elements that may contain caustics onto a grid in order to solve for the gravitational field. The pseudo-particle approach also facilitates the interfacing of our method with existing $N$-body codes.

\item The phase-space elements map from cuboid domains in Lagrangian space to Eulerian space. We show that by subdividing these domains, one can construct a Lagrangian oct-tree based AMR scheme that allows for dynamic adaptive refinement of the elements.

\item In Sections~\ref{sec:mixing_noref} and \ref{sec:mixing_ref}, we demonstrate that in a simple phase-mixing test in a fixed potential dynamic adaptive refinement allows us to conserve energy and represent phase-mixing exactly for the fine-grained distribution function by inserting new vertices. At the same time, new quantities, not accessible for an $N$-body simulation can be calculated. We show, e.g., the growth of the total configuration space volume of the distribution function over time.

\item Next, in Section~\ref{sec:test_1d}, we show that we approach second order convergence with the quadratic elements at fixed vertex number and increasing force resolution in the plane wave collapse test. In the following sections, we provide further one- and two-dimensional tests that demonstrate the high accuracy of our approach and the excellent performance of the Lagrangian phase-space elements compared to $N$-body.

\item Finally, in Section~\ref{sec:cosmo}, we argue that our method is mature
and efficient enough for its application to cosmological structure formation
problems. As an example, we simulate the evolution of the dark matter fluid
with a small-scale cut-off in its initial power spectrum. We explicitly show
that our method solves the artificial fragmentation problem while providing an
unbiased solution for the density field in regions of heavy distortions of
Lagrangian elements.

\end{enumerate}

\noindent Clearly, this new approach will enable us to follow the full fine-grained distribution function throughout gravitational collapse in the formation history of a halo. We expect that knowledge of the distribution of dark matter in six-dimensional phase-space will provide new insights into virialisation as well as relaxation processes in such collisionless self-gravitating systems.

\section*{Acknowledgements}
We thank Tom Abel for the spark to consider adaptive refinement of tetrahedra,
as well as him and St\'ephane Colombi for 
the organisation of various workshops surrounding the Vlasov equation. 
We thank Tom Abel, Ralf Kaehler, Thierry Sousbie, St\'ephane
Colombi, Romain Teyssier, Francesco Miniati, Volker Springel, Simon White,
Devon Powell and, last but not least, Sergei Shandarin for various discussions,
inspirations and/or suggestions along the way. We are indebted to Volker
Springel for his work on {\sc Gadget}, on which we built our implementation.

\noindent O.H. acknowledges support from the Swiss National Science Foundation (SNSF) through the Ambizione fellowship.



\label{lastpage} \end{document}